\documentclass[twocolumn,amsmath,amssymb,prl,10pt,nofootinbib,superscriptaddress]{revtex4}
\usepackage{ graphicx, float,amsmath, amsmath, amssymb}
\usepackage[export]{adjustbox}

\def\be{\begin{equation}}
\def\ee{\end{equation}}
\def\bea{\begin{eqnarray}}
\def\eea{\end{eqnarray}}
\def\bse{\begin{subequations}}
\def\ese{\end{subequations}}

\usepackage[breaklinks, colorlinks, citecolor=blue]{hyperref}
\usepackage[labelsep=period]{caption}
\usepackage{hyperref}
\usepackage[all]{xy}
\usepackage{stmaryrd}
\usepackage{psfrag}
\usepackage{rotating}
\usepackage{epstopdf}
\usepackage{braket}
\usepackage{ulem}

\newcommand{\bit}{\begin{itemize}}
	\newcommand{\eit}{\end{itemize}}
\newcommand{\bd}{\begin{description}}
	\newcommand{\ed}{\end{description}}

\newcommand{\bc}{\begin{center}}
	\newcommand{\ec}{\end{center}}

\newcommand{\bs}{\begin{subequations}}
	\newcommand{\es}{\end{subequations}}

\begin{document}

\title{Phenomenology of Quantum Reduced Loop Gravity in the isotropic cosmological sector}

\author{Emanuele Alesci}%
\affiliation{%
Institute for Gravitation and the Cosmos, Penn State, University Park, PA 16802, U.S.A
}

\author{Aur\'elien Barrau}%
\affiliation{%
Laboratoire de Physique Subatomique et de Cosmologie, Universit\'e Grenoble-Alpes, CNRS/IN2P3\\
53, avenue des Martyrs, 38026 Grenoble cedex, France
}

\author{Gioele Botta}%
\affiliation{%
Faculty of physics, University of Warsaw, Pasteura 5, 02-093 Warsaw, Poland
}

\author{Killian Martineau}%
\affiliation{%
Laboratoire de Physique Subatomique et de Cosmologie, Universit\'e Grenoble-Alpes, CNRS/IN2P3\\
53, avenue des Martyrs, 38026 Grenoble cedex, France
}

\author{Gabriele Stagno}%
\affiliation{%
Sapienza University of Rome, P.le Aldo Moro 5, (00185) Roma, Italy
}
\affiliation{%
Aix Marseille Univ., Univ. de Toulon, CNRS, CPT, UMR 7332, 13288 Marseille, France
}

\date{\today}

\begin{abstract}
Quantum reduced loop gravity is designed to consistently study symmetry reduced systems within the loop quantum gravity framework. In particular, it bridges the gap between the effective cosmological models of loop quantum cosmology and the full theory, addressing the dynamics before the minisuperspace reduction. This mostly preserves the graph structure and SU(2) quantum numbers. In this article, we study the phenomenological consequences of the isotropic sector of the theory, the so-called emergent bouncing universe model. In particular, the parameter space is scanned and we show that the number of inflationary e-folds is almost always higher than the observational lower bound. We also compute the primordial tensor power spectrum and study its sensitivity upon the fundamental parameters used in the model.
\end{abstract}

\maketitle

\section{Introduction}

	The Higgs boson discovery \cite{Aad:2012tfa} and the direct observation of gravitational waves \cite{Abbott:2016blz} have strengthened the reliability of well corroborated theories: the standard model of particle physics (based on quantum field theory) on the one hand, and general relativity (GR) on the other hand. Beside these recent observations, the long-standing issue of quantizing gravity still calls for a solution. All physical theories must make contact with experiments or observations and this often constitutes one of the main difficulties for quantum gravity. Loop quantum gravity \cite{Ashtekar:2004eh,Rovelli:2004tv,Thiemann:2007zz} (LQG) is a consistent attempt in this direction, as witnessed by the recent effort on dealing with the black hole quantum dynamics (both within the canonical \cite{Alesci:2018loi, Ashtekar:2018cay} and covariant formulations \cite{Christodoulou:2016vny}), together with the prediction of the big bang singularity resolution \cite{Bojowald:2001xe, Ashtekar:2006uz,Ashtekar:2006wn} and the power spectrum calculation \cite{Agullo:2013ai,Ashtekar:2015dja,Agullo:2017eyh} made possible by loop quantum cosmology (LQC).\\
	 
	 This article is about the observable consequences of LQG in cosmology, when the full theory structure is taken into account. This can be done using a suitable gauge fixed version of the theory called quantum reduced loop gravity (QRLG) \cite{Alesci:2012md, Alesci:2013xd,  Alesci:2014uha, Alesci:2014rra, Alesci:2015nja, Alesci:2016gub}. Differences between LQC and QRLG are both in the philosophy and  the methodology. The former is a LQG-inspired, polymerlike \cite{Agullo:2016tjh,Corichi:2007tf} quantization of a classically symmetry reduced system, while the latter is a subsector of LQG adapted to the symmetry of the system one is interested in. In the two approaches, quantization and symmetry reduction are in reverse order: LQC quantizes a classical reduced system, QRLG selects a symmetric subsector from the full quantum theory. If LQC can be seen as the simplest and most straightforward application of LQG ideas, starting from the beginning with less degrees of freedom to quantize, from the QRLG perspective it can be trusted as a first order quantum correction to the classical dynamics, since relevant structures of LQG are lost and have to be ``injected" in the process. On the contrary, QRLG retains all the features of the full theory and, moreover, does indeed recover LQC at first order \cite{Alesci:2016rmn,Alesci:2017kzc}.\\ 
     
      In this article we extend the study of QRLG addressing inflation and discussing features of the power spectrum for cosmological perturbations. In isotropic QRLG, the Friedman Lemaitre Robertson Walker (FLRW) background is replaced by an emergent bouncing universe \cite{Alesci:2016xqa}. Here we focus on observable signatures of this scenario and compare them to the ones provided by LQC. As shown in \cite{Alesci:2016rmn,Alesci:2017kzc}, the corrections are subleading only up to the (first) bounce -- when going backward in time -- and, for earlier times, they grow and lead to a complete different dynamics. Thus, the observational consequences of the QRLG scenario have to be studied as they may differ from LQC ones. Before introducing our model, we briefly review LQG in order to make possible the understanding of our results also to the reader unfamiliar with the full theory.\\
     
     LQG is a background free, nonperturbative Hamiltonian quantization of gravity whose starting point is the 3+1 foliation of the GR first order tetradic formulation. It is a modern canonical quantization that takes advantage of a new set of phase space variables -- the Ashtekar variables \cite{PhysRevLett.57.2244} -- in order to cast the classical theory in a form close to the one of a local SU(2) gauge theory. 
The Ashtekar variables $A_{a}^i(x;t)\,,E^a_i(x;t)$ are an $su(2)$ connection and a (densitized) triad field, which are canonically conjugate, $\{A_{a}^i(x),E^b_j(y)\}=8\pi G\gamma\,\delta^b_a\delta^i_j\delta^3(x-y)$, and read 
$A_{a}^i:=\omega^i_a +\gamma K_a^i\,,E^a_i:=\frac{1}{2}\epsilon_{ijk}\epsilon^{abc}e_b^je_c^k\,,$
where $i,j,k$ are $su(2)$ algebra indices, $a,b,c$ space ones, $\omega^i_a$ is the spin-connection compatible with the triad $e^b_j$, $K_a^i$ is the (mixed triadic projection of the) extrinsic curvature tensor and $\gamma$ is a parameter that enters this formulation of GR. This so-called Barbero-Immirzi parameter $\gamma$ is expected to have a value close to $0.24$ if one considers the black hole entropy calculation \cite{Meissner:2004ju}. It enters in the spectrum of the geometrical operators like area and volume, but does not change the classical equations of motion, {\it i.e.} Einstein's equations. Like all gauge theories, GR is a constrained system, more specifically, a \textit{totally} constrained one, as its Hamiltonian vanishes on physical trajectories.  Written in Ashtekar variables, it turns out to be encoded in three constraints generating SU(2) gauge transformations (the Gauss constraint), spatial diffeomorphisms (the Diffeomorphism constraint) and time reparametrization (the Hamiltonian constraint). \\

 Quantization starts using a ``technology" borrowed from lattice gauge theories in order to provide a (background-independent) smearing of the canonical algebra generated by $A_{a}^i(x;t)\,$ and $E^a_i(x;t)$, leading to the holonomy-flux algebra. The Ashtekar connection $A_a^i(x)$ is replaced by its holonomy $h_{l}[A]$ along arbitrary paths $l$- and the densitized triad $E^a_i(x)$ is replaced by its flux $E_i(S)$ across a surface $S$. Quantization follows implementing the (unique \cite{Lewandowski:2005jk}) quantum representation of the holonomy-flux algebra and computing the kernel of all the quantum operator-promoted constraints of the theory, according to Dirac's procedure \cite{DIRAC} for constrained systems. Solving the Gauss and Diffeomorphism constraints leads to the definition of a Hilbert space with states $|\Gamma,j,i\rangle$. Those states are labeled by graphs $\Gamma$ given by links associated to the holonomies (dual to the surfaces used for defining fluxes) and nodes. Links are colored by spins $j$, {\it i.e.} by representations of $SU(2)$, and nodes by intertwiners $i$, {\it i.e.} $SU(2)$ invariant tensors.  Geometric quantities can be turned in Hermitian operators and it turns out that they have a discrete spectrum \cite{Rovelli:1994ge}. The area operator has a spectrum with a minimal nonvanishing eigenvalue $\Delta= 4\sqrt{3}\pi G\,\gamma\, l_P^2$ proportional to the Barbero-Immirzi parameter $\gamma$ and the square of the Planck length $l_P:=\sqrt{\hbar G/c^3}$.
The picture provided by LQG is clear and beautiful: quantum gravity appears as a quantum theory of geometry, in which the spacetime continuum disappears leaving place to a relational net of fuzzy quanta of space.\\

Beside these achievements, problems arise when addressing the Hamiltonian constraint.  Only trivial and formal solutions \cite{Alesci:2011ia} are indeed known and a complete characterization of the full spectrum  is still missing. A retrospective look at this difficulty is not so discouraging: after all, the general solution to the analogue classical problem, {\it i.e.} Einstein's equations, is still unknown too, but this has not prevented GR to become a powerful tool for gravity. During the past years several paths to overcome the issue of quantum dynamics have been followed, both implementing different reformulations, {\it e.g.} using spinfoam models \cite{Perez:2012wv}, and/or addressing the dynamics of \textit{ symmetric } sectors of the full theory. The pioneering spin-off of LQG that follows this last direction is the "minisuperspace" quantization of spacetimes pursued by LQC.\\

Calculating the Ashtekar variables for a chosen spacetime, LQC follows a polymerlike quantization that \textit{mimics} the one pursued by LQG and provides the quantum dynamics for symmetry reduced models at the classical level, such as FLRW and Bianchi spacetimes \cite{Ashtekar:2011ni}. The resolution of the cosmological singularity comes out naturally, replacing the Big bang scenario by a nonsingular bouncing universe.  Looking forward in time, there is a contracting phase which ends when the density and the curvature reach near-Planckian values, then a bounce happens and an expanding phase follows (the late-time behavior is exactly as in GR). The singularity is resolved because even though zero is in the spectrum of the volume operator, it is never dynamically reached. Despite this remarkable result, one should look at traditional LQC as a first attempt in applying LQG ideas to the simplest class of gravitational symmetry-reduced systems. The limits of this approach are mainly due to the fact that the quantization is performed only after a classical symmetry reduction and this does not prevent ambiguities in the corresponding quantum theory (see {\it e.g.} \cite{Gupt:2011jh}). Working only with few degrees of freedom, LQC needs to import from the full theory both a graph structure and a minimum value for physical areas in order to regularize the symmetry-reduced Hamiltonian operator.\\

QRLG is a program that attempts to implement a dynamical reduction of the full theory to a given symmetry-reduced setting, {\it i.e.} first quantizes and then reduces. This is achieved in several steps: one begins by implementing a gauge fixing at the quantum level (defining a gauge-fixed kinematical Hilbert space, called the reduced  Hilbert space $\mathcal{H}^R$) and then one uses coherent states peaked on symmetric spaces over which one evaluates the operator version of a new set of constraints that preserve the gauge (built according to the gauge unfixing procedure \cite{Mitra:1989fg, MITRA1990137, Neto:2009rm}).\\

 In the cosmological setting of the FLRW geometry (and Bianchi models), this reduced space is selected by (partially) gauge fixing the SU(2) and the Diffeomorphism gauge of the full theory to diagonal metrics and triads. Only a small class of spatial diffeomorphisms are still compatible with this choice (called \textit{reduced diffeomorphisms}), leading to the result that at the quantum level only cuboidal graphs (colored with $U(1)$ representation numbers) are allowed, {\it i.e.} the ones with links parallel to the fiducial triad field. Computing expectation values of the (gauge preserving part of the) LQG Hamiltonian constraint, QRLG effective Hamiltonians for the FLRW and Bianchi I cases can be explicitly obtained \cite{Alesci:2017kzc}. They depend on the choice of coherent states used to define the symmetry-reduced sectors.\\
 
 Importantly, the much discussed $\mu_0$ or $\bar{\mu}$ LQC regularization schemes appear in QRLG as particular choices of coherent states. QRLG allows to reproduce LQC schemes and to generalize them  \cite{Alesci:2016xqa,Alesci:2017kzc} with the so-called {\itshape statistical} regularization. This is  based on ensembles of coherent states peaked on homogeneous phase space points defining macrostates.  Every homogenous coherent state at a fixed graph represents a given cosmological \textit{macrostate} and statistical superposition of graphs can be considered. To the same macrostate (labeled by $(a,\dot{a})\,,$ for FLRW) corresponds several coherent \textit{microstates} labeled by different quantum numbers and graphs. For each given probability distribution counting the occurence of microstates associated to a fixed macrostate, an effective Hamiltonian can be computed taking the expectation value of the Hamiltonian operator over the chosen ensemble, as done in the aforementioned references where Gaussian ensembles were chosen.\\
 
  All the computed QRLG effective Hamiltonians bring corrections to the LQC ones \textit{that are subleading only much after the Big Bounce}. For the FLRW case, at earlier times, the Universe oscillates and eventually reaches a stationary phase of constant finite volume (the meaning of the ``volume of the Universe" will be discussed later on). Looking forward in time, a Planckian universe \textit{emerges} from the infinite past. It is stationary until a transient phase is reached and, after few bounces, the dynamics matches the LQC's one from the (last) Big bounce all the way to the far future. This emergent behavior is a peculiar property of the isotropic sector and exploring its observational consequences constitutes the main goal we address in the rest of the paper. As far as perturbations are concerned, we use here the usual formalism and we apply only QRLG correction to the background. This is a heavy hypothesis.

In the next section, the effective quantum background is described. Then, the corresponding basic features are investigated. At the background level, the duration of inflation is calculated for most of the parameter space. Regarding perturbations, the tensor power spectra are computed and scalar ones discussed. Finally, the effects of the inflaton field mass are considered

\section{Effective quantum backgrounds}

\subsection{FLRW loop quantum cosmology}

We briefly review here the quantization of the (spatially flat) FLRW spacetime as pursued by LQC, focusing on the effective equations of motion it provides. Starting from the FLRW line element 

\begin{equation}
ds^2=-dt^2+a(t)^2\,\delta_{ij}\,e^i_a e^j_b\,dx^adx^b\,,\label{FLRW_ds}
\end{equation}

where $e^i_a:=\delta^i_a\,,$ is a fiducial triad field in Cartesian comoving coordinates $(t,x,y,z)$ and  $a(t)\,$ is the scale factor. The associated Ashtekar variables are computed in order to write the FLRW Hamiltonian provided by GR in terms of them. To this aim, a fiducial cell of coordinate volume $V_0$ is introduced\footnote{This regulator can be removed at the end: the usual Friedmann equation of motions as well as the effective LQC ones \eqref{LQC_1Fried} and \eqref{LQC_2Fried} do not depend on it. Note that for the QRLG model, this is not the case and the initial physical volume of the Universe turns out to be a parameter  that has to be constrained by data -- this will be discussed later.} so as to avoid spurious divergences due to the open topology this geometry is (here implicitly) endowed with. Now, thanks to the symmetries of \eqref{FLRW_ds}, the spin connection is vanishing, the extrinsic curvature tensor is proportional to the time derivative of the scale factor and the Ashtekar variables assume the simple expressions

\begin{equation}
A_{a}^{i}(t)=c(t)\delta_{a}^{i} V_{0}^{-1/3}\,, \ \ \ E_{i}^{a}(t)=p(t)\delta_{i}^{a} V_{0}^{-2/3}\,, \label{lqcvar1}
\end{equation}

where 

\begin{equation}
c:=V_{0}^{1/3}\gamma \dot{a}\,, \quad p:=a^{2}V_{0}^{2/3}\quad \mbox{and}\quad \{c,p\}=\frac{8\pi \gamma}{3}\,,\label{cpvariables}
\end{equation}

and the FLRW Hamiltonian constraint reads 

\begin{equation}
\mathcal{H}=-\frac{3}{8\pi\gamma^2}\,\sqrt{p}\,c^2\ =0~,\label{H_classical}
\end{equation}

as one can easily check computing the associated Hamilton equations of motion. The usual Friedmann equations are obtained from them once $a$ and $\dot{a}$ are inverted from \eqref{cpvariables} and the appropriate matter content is added.
The next step consists in switching from this classical model to its quantum version by implementing a suitable quantum representation of the canonical variables \eqref{lqcvar1}: LQC mimics LQG by computing holonomies from the Ashtekar connection and fluxes from the triads. Thanks to the symmetry of the FLRW spacetime, one can consider only holonomies $h_{\mu}(c)$ along edges of the fiducial cell and fluxes $E(S)$ across faces $S$ of $V_{0}$:

\begin{equation}
h_{\mu}(c):=e^{i\mu c/2}\,, \ \ \ \ E(S):=p\,,\label{LQC_qvariables} 
\end{equation}

where $\mu$ is the ratio between the coordinate length of a path parallel to an edge of the fiducial cell and the length of the edge itself. \\

Once the classical constraint\footnote{This is the only constraint one remains with, as the Gauss and Diffeomorphism ones are trivially fulfilled thanks to the symmetry reduction.} \eqref{H_classical} is written in terms of \eqref{LQC_qvariables}, it can be promoted to be a quantum operator after a regularization for the final chosen expression. LQC takes again inspiration from LQG, where geometry is discretized and areas exhibit of a minimum area gap $\Delta$ (on regularizations in LQC see \cite{Alesci:2017kzc}). This feature also arises in this reduced setting through a regularization (so-called ``improved" \cite{Ashtekar:2006wn}) achieved by promoting the $\mu$ parameter entering in the holonomies to a be a \textit{function} $\bar{\mu}:=\bar{\mu}(p)\,$.
Computing the expectation value of the resulting Hamiltonian operator over coherent states peaked in the classical phase space of the FLRW geometry $(c,p)$, one obtains an effective Hamiltonian \cite{Taveras:2008ke,Ashtekar:2015iza} $H^{LQC}$ that  we can simply introduce by the following ``rule", also called "polymer" substitution in \eqref{H_classical},

\begin{equation}
c\rightarrow\frac{\sin(\bar{\mu}c)}{\bar{\mu}}\quad\mbox{where}\quad\bar{\mu}:=\sqrt{\frac{\Delta}{p}}\,,\label{polysubstitution}
\end{equation} 

which leads to the following effective LQC Hamiltonian for the geometric sector:

\begin{equation}
\mathcal{H}^{LQC}_{grav}:=-\frac{3}{8\pi\gamma^2}\,\sqrt{p}\,\frac{\sin^2(\bar{\mu}c)}{\bar{\mu}^2}\,.\label{LQC_H_eff_geo}
\end{equation}

In the quantum theory, the basic variables are \eqref{LQC_qvariables} and there exist no quantum operator $\hat{c}$ corresponding to $c$. The polymer substitution can be considered as a trigonometric approximation of $\hat{c}$, when written as the derivative of $h_{\mu}$ evaluated in $\mu=0$, 

\begin{equation}
c=\frac{2}{i} \left. \frac{d}{d\mu}h(c)\right| _{\mu=0}\approx\frac{2}{i}\frac{h_{2\mu}(c)-h_{-2\mu}(c)}{2\mu}\,,
\end{equation}

followed by the replacement $\mu\rightarrow\bar{\mu}$ that defines the specific regularization adopted by LQC.\\

When the FLRW geometry is sourced by a (minimally coupled) massless scalar field, one adds to the effective Hamiltonian \eqref{LQC_H_eff_geo} its kinetic contribution, {\it i.e.} $\mathcal{H}_{\phi}:=P_{\phi}^2/(4\pi\gamma v)$ where $P_{\phi}$ is the momentum conjugate to the field $\phi(t)\,,\{\phi,P_{\phi}\}=1\,,$ and the complete Hamiltonian reads

\begin{equation}
\mathcal{H}^{LQC}_{grav+\phi}:=-\frac{3v}{4\Delta\gamma}\,\sin^2(b\sqrt{\Delta})+\mathcal{H}_{\phi},\label{H_LQC}
\end{equation}

after the change of variables $(c,p)\rightarrow(b,v)$, where

\begin{equation}
 b:=\frac{c}{p^{1/2}}\,,\quad v:=\frac{p^{3/2}}{2\pi\gamma}
\,, \qquad\left\lbrace \frac{b}{\sqrt{2}},\frac{v}{\sqrt{2}} \right\rbrace=1\,.\label{bev} 
\end{equation}

Finally, the effective dynamics is obtained through the Hamilton equations of motion:

\begin{equation}
\dot{Q}_i=\left\lbrace Q_i,\mathcal{H}^{LQC}_{grav+\phi}\right\rbrace \,,\quad\dot{P}_i=\left\lbrace P_i,\mathcal{H}^{LQC}_{grav+\phi}\right\rbrace,
\end{equation}

where

\begin{equation}
Q_i:=\left( \frac{b}{\sqrt{2}},\phi\right) \quad\mbox{and}\quad P_i:=\left( \frac{v}{\sqrt{2}},P_{\phi} \right) \label{Effective_Hamiltoneqq},
\end{equation}

and the Poisson brackets are defined on the whole phase space $(b,v)\times(\phi,P_{\phi})$:

\begin{equation}
\{\,,\,\}:=\sum_{i}\frac{\partial}{\partial Q_i}\frac{\partial}{\partial P_i}-\frac{\partial}{\partial P_i}\frac{\partial}{\partial Q_i},
\end{equation}

 giving
 
\begin{eqnarray}
\frac{\dot{a}^2}{a^2}&=&\frac{8\pi}{3}\rho_m\left(1-\frac{\rho_m}{\rho_{crit}}\right);\label{LQC_1Fried}\\
\frac{\ddot{a}}{a}-\frac{\dot{a}^2}{a^2}&=&-8\pi\rho_m\left(1-2\frac{\rho_m}{\rho_{crit}}\right)\,,\label{LQC_2Fried}
\end{eqnarray}

where $\rho_m:=P^2_{\phi}/(8\pi^2\gamma^2 v^2)$ is the scalar field energy density and $\rho_{crit}=3/(8\pi\gamma^2 \Delta)$ is the critical energy density (depending on the LQG minimum area gap $\Delta$) at which the Universe undergoes a bounce. In fact, one can immediately see that $\rho_m=\rho_{crit}$ in \eqref{LQC_1Fried} and \eqref{LQC_2Fried} corresponds to a stationary point. The ``repulsive force" encoded in the $\rho^2_m$ correction to the Friedmann equation reacts to classical gravity when the energy density reaches a near-Planckian value and the singularity is tamed. The discretness of space predicted by LQG, and imported in LQC, leads to the singularity resolution. In this framework, the bounce happens to occur when a Planckian value of the \textit{energy density} is reached, regardless of the volume of the Universe -- or of the ``fundamental cell" -- that can be anything, as \eqref{LQC_1Fried} and \eqref{LQC_2Fried} depend only on the scale factor (and the chosen value for $P_{\phi}$).\\

The Big bounce scenario is a robust prediction of LQC, as witnessed by its persistence when nonvanishing potentials are added \cite{Ashtekar:2011ni,Zhu:2017jew}. This remains true with curvature \cite{Singh:2003au} and with a cosmological constant \cite{Pawlowski:2011zf}. In the following we will focus on inflation and consider the case of a \textit{massive} scalar field with a quadratic potential. The LQC dynamics associated to the corresponding effective Hamiltonian,
\begin{equation}
\mathcal{H}^{LQC}_{grav+\phi^2}:=\mathcal{H}^{LQC}_{grav}+\frac{P^2_{\phi}}{2V}+V \frac{m^2\phi^2}{2}\,,\label{H_LQCpotential}
\end{equation}
where $V:=2\pi\gamma v\,,$ gets (qualitatively) unchanged until the beginning of the slow-roll inflationary phase generated by the massive field.

\subsection{QRLG emergent-bouncing universe}\label{subsection_emergent}

The approach pursued by QRLG greatly simplifies the LQG computational task, especially when addressing isotropic cosmology. In particular, one can easily calculate the effective dynamics for a quantum corrected FLRW universe, evaluating the expectation value of the LQG Hamiltonian constraint over a mixture of coherent states based on cubical graphs with different numbers of nodes $N$, and peaked on the classical FLRW phase space coordinates.
 
 The model provided by QRLG has the same symplectic structure than LQC, defined by \eqref{bev}, the only difference being in the  Hamiltonian and the effective dynamics. We report here its final expression computed within the so-called volume counting statistical regularization scheme \cite{Alesci:2017kzc}, where a Gaussian distribution of coherent states centered on $N=V\tilde{\Delta}^{-3/2}$ is chosen:

\begin{eqnarray}
&& \mathcal{H}^{QRLG}_{full}(V,b)= -\frac{3}{8\pi\gamma^2}V^{1/3} \\ \nonumber
&& \times \frac{\int_1^{2V\tilde{\Delta}^{-3/2}} \,e^{-\frac{(N-V\tilde{\Delta}^{-3/2})^2}{V\tilde{\Delta}^{-3/2}}}\,N^{2/3} \sin^{2}\left(\frac{bV^{1/3}}{N^{1/3}}\right)\,dN}{\int_1^{2V\tilde{\Delta}^{-3/2}} \,e^{-\frac{(N-V\tilde{\Delta}^{-3/2})^2}{V\tilde{\Delta}^{-3/2}}}\,dN}\,, \label{ACTUALhamvb}
\end{eqnarray}

where $V$ is a physical volume $(V=a^3V_0)$ and $\tilde{\Delta}$ is related to the LQG area gap by $\tilde{\Delta}:= 2^{2/3}\sqrt{3}\,\Delta$. (It should be noticed that there are two different $\Delta$ parameters for two reasons. In QRLG, the ``reduced flux" operator -- from which the QRLG area operator is built -- turns out to have eigenvalues that are proportional to $m$ and not to $\sqrt{j(j+1)}$, like in LQG, thus only for $j\gg 1$ do the two definitions match. Beside, a further deviation from the ``standard" $\Delta$ comes from the actual density matrix chosen to regularize the effective Hamiltonian within the statistical regularization scheme.)
The Hamiltonian defines the geometrical sector of the model. Adding the usual kinetic contribution for a massless scalar field $\phi$ and considering the first order contribution to the saddle point approximation for $V\gg 1$, one is led to the following approximated Hamiltonian that describes geometry and matter:

\begin{eqnarray}
\mathcal{H}^{QRLG}_{1ord}+\mathcal{H}_{\phi}&=&-\frac{3v}{4\tilde{\Delta}\gamma}\,\sin^2(b\sqrt{\tilde{\Delta}})+\frac{P_{\phi}^2}{4\pi\gamma v} \nonumber\\
&-& \frac{ b^2{\tilde{\Delta}}^{3/2}}{48\pi\gamma^2}\cos(2b\sqrt{\tilde{\Delta}}) \\ \nonumber
&+&\frac{\sqrt{\tilde{\Delta}}}{48\pi\gamma^2 }\sin^2(b\sqrt{\tilde{\Delta}})\,,\label{ciao2}
\end{eqnarray}

and already captures the relevant features of the model, allowing analytical considerations for the qualitative behavior of the associated dynamics. In the first line, one immediately recognizes an LQC-like contribution (which up to the area gap $\tilde{\Delta}$ redefinition, exactly coincides with the expression \eqref{H_LQC}) while the second and third lines correspond to the (first order) QRLG corrections: those are subleading in the semiclassical regime $b/v\ll 1$, where LQC and QRLG dynamics match, but become leading orders in the deep quantum epoch, giving a very different dynamics, as discussed below and shown in the upper panel of Fig. \ref{v vs t}.\\

Using $\mathcal{H}^{QRLG}_{1ord}+\mathcal{H}_{\phi}$ and neglecting terms that are subdominant in a $1/v$ expansion\footnote{In writing eq.(\ref{F5}) the last term in the rhs of (\ref{ciao2}) has been neglected but all the numerical studies have been done keeping also that contribution.}, the following modified Friedmann equation are found

\begin{eqnarray}
\frac{\dot{a}^{2}}{a^{2}}= && \left(\frac{8 \pi}{3} \rho_{\rm m}  +\frac{\rho_{\rm g}}{\gamma^2} \right) (1-2\Omega_{\rm g})^{-1} \\ \nonumber
&& \times \left(1-\frac{\Omega_{\rm m}-\Omega_{\rm g}}{1-2\Omega_{\rm g}}\right)\,,\label{F5}
\end{eqnarray} 

\begin{eqnarray}
\frac{\ddot{a}}{a}-\frac{\dot{a}^2}{a^2}= &-& \left(\frac{3}{\tilde{\Delta}\gamma^2}\sin^2\left( b\sqrt{\tilde{\Delta}}\right) +4\pi \rho_{\rm m}\right) \\ \nonumber &\times &\left(1-2\sin^2\left( b\sqrt{\tilde{\Delta}}\right) \right),\label{F6}
\end{eqnarray}

where 

\begin{eqnarray}
&&\rho_{\rm g}:=-\frac{b^2{\tilde{\Delta}}^{3/2}}{18 V} ~, ~~  \bar{\rho}_{\rm cr}:=-\frac{1}{\tilde{\Delta}} ~, \\ \nonumber
&& \rho_{ m}:=\frac{P_{\phi}^{2}}{2V^{2}} ~, ~~ \rho_{cr}:=\frac{3}{8\pi\gamma^{2}\tilde{\Delta}} ~, \\ \nonumber 
&&\Omega_{\rm g}:=\frac{\rho_{\rm g}}{\bar{\rho}_{\rm cr}} ~, ~~ \Omega_{\rm m}:=\frac{\rho_{\rm m}}{\rho_{\rm cr}}\,.\label{V1}
\end{eqnarray}

The quantity $\rho_{\rm g}$ is interpreted as a pure quantum gravitational (negative) energy density, vanishing for $\tilde{\Delta}\rightarrow 0\,,$ and $\bar{\rho}_{cr}$ is the critical energy density at which an empty universe ($\rho_m=0$) would undergo a bounce. The former is a key quantity of the model since for $\rho_{\rm g}\rightarrow 0$ Eqs. \eqref{F5} and \eqref{F6} give back the LQC effective dynamics, {\it i.e.} \eqref{LQC_1Fried} and \eqref{LQC_2Fried} with an area gap $\tilde{\Delta}$. Two conditions lead to a stationary point:

\begin{equation}
\Omega_{\rm g}+\Omega_{\rm m}=1\,, \ \ \ \Omega_{\rm g}=\Omega_{\rm m}\,. \label{cond}
\end{equation}

The first is similar to what happens in LQC, since for $\Omega_{\rm g}\rightarrow 0$ it gives $\Omega_{\rm m}=1\,.$ When the sum of the ratio between $\rho_{\rm g}/\bar{\rho}_{cr}$ and $\rho_{ m}/\rho_{cr}$ is equal to $1$, the Universe bounces reaching a local minimum of the volume. The second condition is the main novelty brought by QRLG: when quantum gravity effects compensate the evolution driven by the matter content, maxima are reached and, going back in time, the LQC prebounce dynamics is replaced by oscillations with decreasing amplitudes (see upper and middle panels of Fig. \ref{v vs t}).\\ 

The picture provided by the isotropic sector of QRLG is an asymmetric scenario of the primordial universe: the Universe emerges from the infinite past with a finite Planckian volume and eventually undergoes a transient phase during which expanding and contracting phases succeed until the geometric energy density gets enough diluted (as $b$ decreases) to leave the Universe expanding forever according to the classical dynamics. As we will show later, this behavior, discovered for a massless scalar field, is qualitatively unchanged for a massive scalar field.

\section{Background dynamics: basic features}

The phenomenology of QRLG is a tricky task. There is indeed a fundamental tension between the basis of QRLG and usual cosmology. Friedmann equations are invariant under a rescaling of the scale factor. There is no preferred length scale in cosmology. If the curvature is null or negative, the size of the Universe is infinite at all times. This is why, in usual LQC (see {\it e.g.} \cite{Ashtekar:2015dja,lqc9}), the bounce is driven by {\it density} effects (together with the shear). However, in QRLG, there is a physical scale associated with the fundamentally discrete structure of space. This does not mean that QRLG is inconsistent: the other way around, this is expected at the quantum geometrical level. It means, as advocated {\it e.g.} by Bojowald \cite{Bojowald:2015iga}, that quantum cosmology might not be about quantizing the scale factor and its conjugate variable (say the Hubble parameter) but about the dynamics of elementary and identical cells of space. What is usually referred to as the ``volume of the Universe" should probably be actually understood as the volume of an elementary patch. Although there is therefore no logical inconsistency, several issues about making concrete predictions in this framework remain open due to the nontrivial transition between the effective quantum description and the classical regime.\\ 

The background evolution is driven by the full Hamiltonian given in the previous section but the matter content is now chosen to be a massive scalar field with a mass $m=1.21\times10^{-6}$ (unless otherwise stated Planck units are now used). Although slightly disfavored by recent data \cite{Ade:2015lrj}, this is a standard choice in cosmology  which is also frequently done in LQC, in order to make comparisons between models easier. The following results may depend on the field mass and this will be addressed later in the text.\\

The status of initial conditions in QRLG remains a complicated question, as there are no \textit{a priori} preferred probability density functions for the different parameters in the quantum regime (we shall address this point again in the final discussion). However, the late classical universe is described by a large $v$ value and a small $b$ value. Taking this into account, the background evolution can be explicitly computed, as a first step, evolving the state backward in time. We used a numerical simulation with initial conditions set in the classical phase as in previous works in QRLG \cite{Alesci:2016xqa},

\begin{eqnarray}
&& v_{class} = 100 000 ~,\\ \nonumber
&& b_{class} = 0.0005 ~,\\ \nonumber
&& P_{\phi ,class} = 88 ~.
\label{IC classical}
\end{eqnarray}

The initial condition on the last parameter $\phi_{class} $ is obtained thanks to the Hamiltonian constraint $\mathcal{H}=0$.\\ 

\begin{figure}[!tbp]
  \begin{minipage}[t]{0.4\textwidth}
  \hspace{1.3 cm}
\includegraphics[scale=0.55]{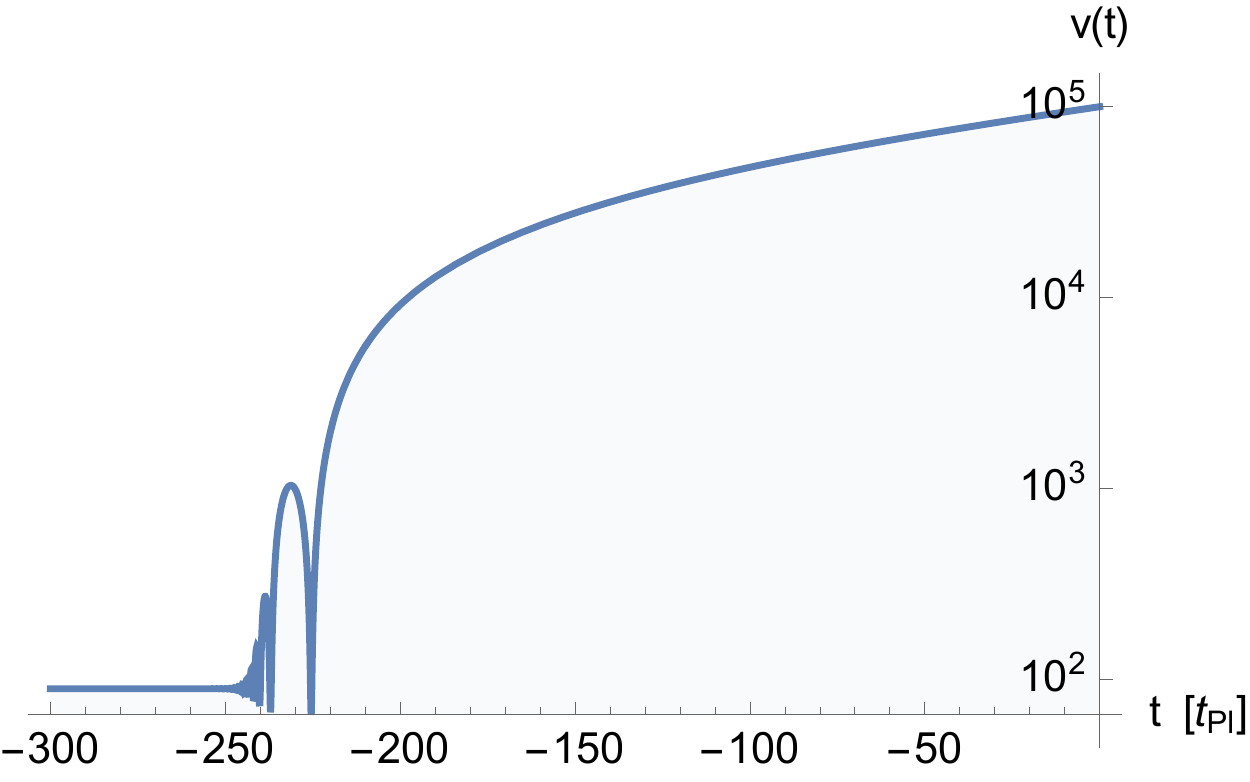}
\label{v backward}
  \end{minipage}
  \vspace{0.2 cm}
  \begin{minipage}[t]{0.4\textwidth}
\includegraphics[scale=0.55]{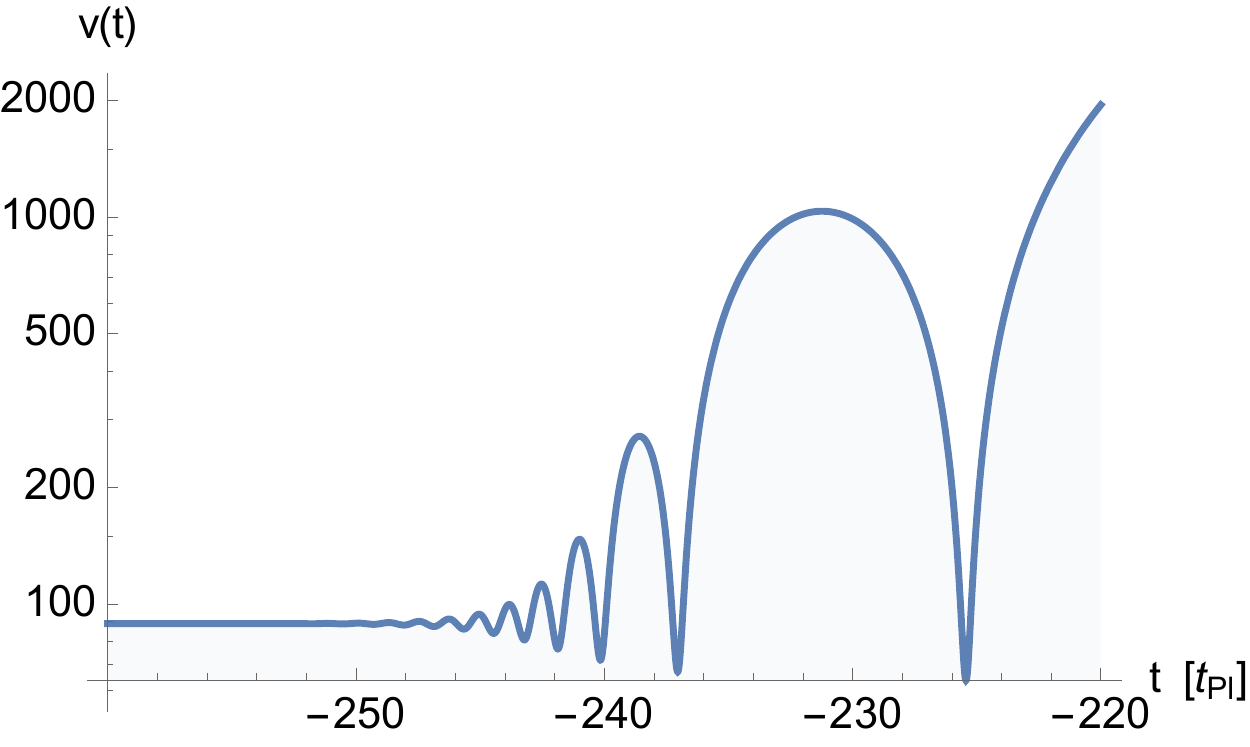}
\label{v backward zoom}
  \end{minipage}
    \begin{minipage}[t]{0.4\textwidth}
\includegraphics[scale=0.375]{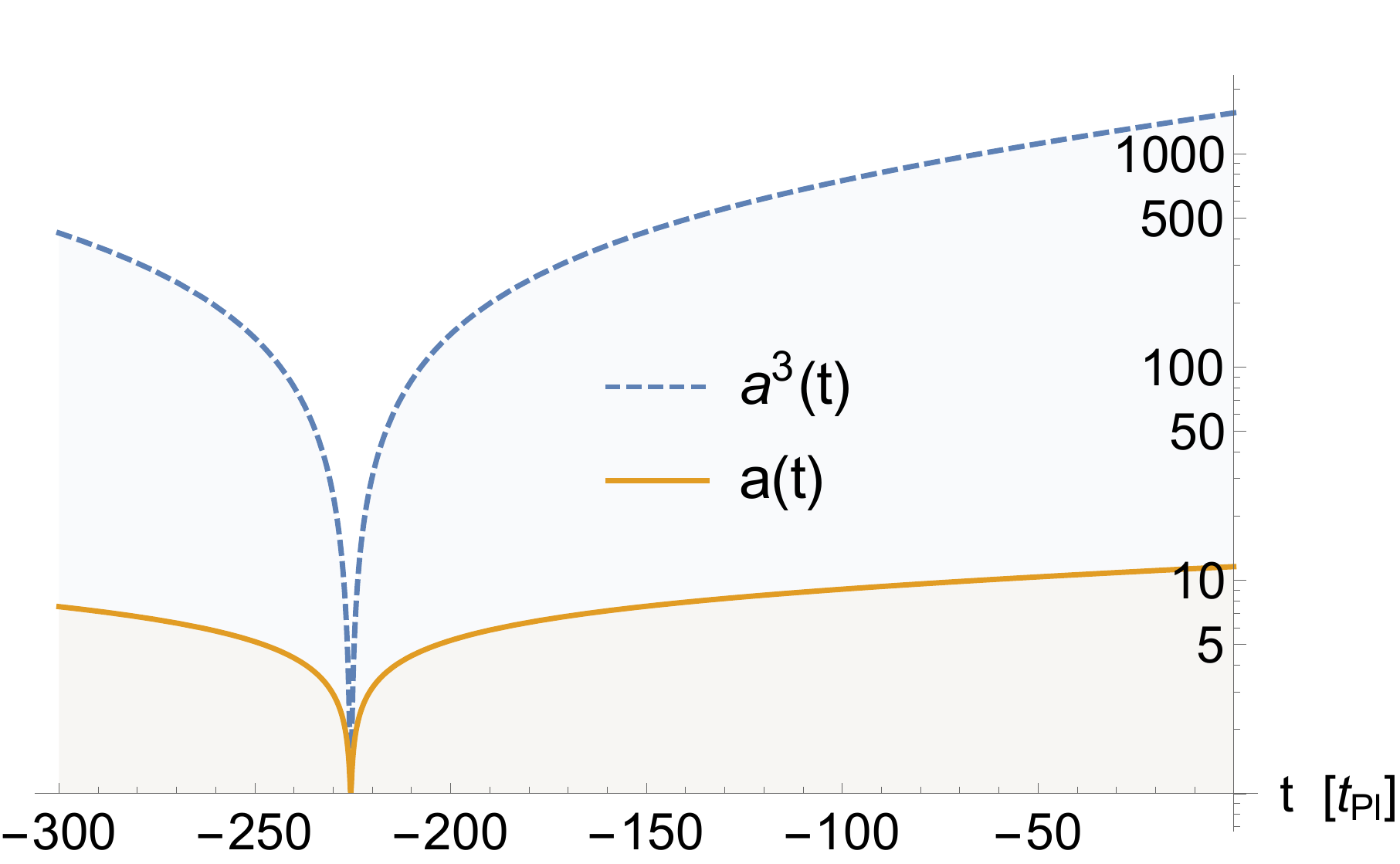}
\label{lqc bounce}
  \end{minipage}
   \caption{\underline{Upper panel:} $v(t)$ backward evolution, starting from $t=0$. \underline{Middle panel:} Zoom on the emergent phase. \underline{Lower panel:} The LQC scale factor backward evolution obtained starting from the same initial conditions at $t=0$.} 
   \label{v vs t}
\end{figure}

This backward evolution leads, as explained before, to bounces of decreasing amplitudes that converge to a (quasi)static phase, as shown on Fig. \ref{v vs t}. This is an interesting mixture between emergent and bouncing models. It should be noticed that this dynamics is basically the same for different sets of initial conditions in the classical phase, as long as $v_{class}$ is large and $b_{class}$ is small. The lower panel of Fig. \ref{v vs t} represents the scale factor evolution in LQC, obtained from the same classical initial conditions at $t=0$. It can be observed that the QRLG evolution is indeed the same as the LQC one up to the bounce (when evolving backward in time) but the LQC dynamics then leads to a classical contracting branch.\\ 

The values of $\left\lbrace v,b,\phi,P_{\phi} \right\rbrace$ in the static phase, \textit{i.e} at $t=-300$ in our simulation, obtained from this backward evolution are then used as preferred initial conditions to perform simulations forward in time. In the following, those new initial conditions are denoted as $\left\lbrace v_\text{in},b_\text{in},\phi_\text{in},P_{\phi,\text{in}} \right\rbrace$. This procedure is helpful for the gravitational variables $\left\lbrace v,b \right\rbrace$, as their values in the quantum (quasi)static regime is set by  physical arguments requiring a correct classical behavior. However, this does not constrain the matter content: the value of the scalar field and its momentum are still free. The consequences of the possible choices for initial field conditions on the different observables will be studied later.\\ 



\begin{figure}[!tbp]
  \centering
  \begin{minipage}[t]{0.4\textwidth}
  \hspace{0.5 cm}
\includegraphics[scale=0.55]{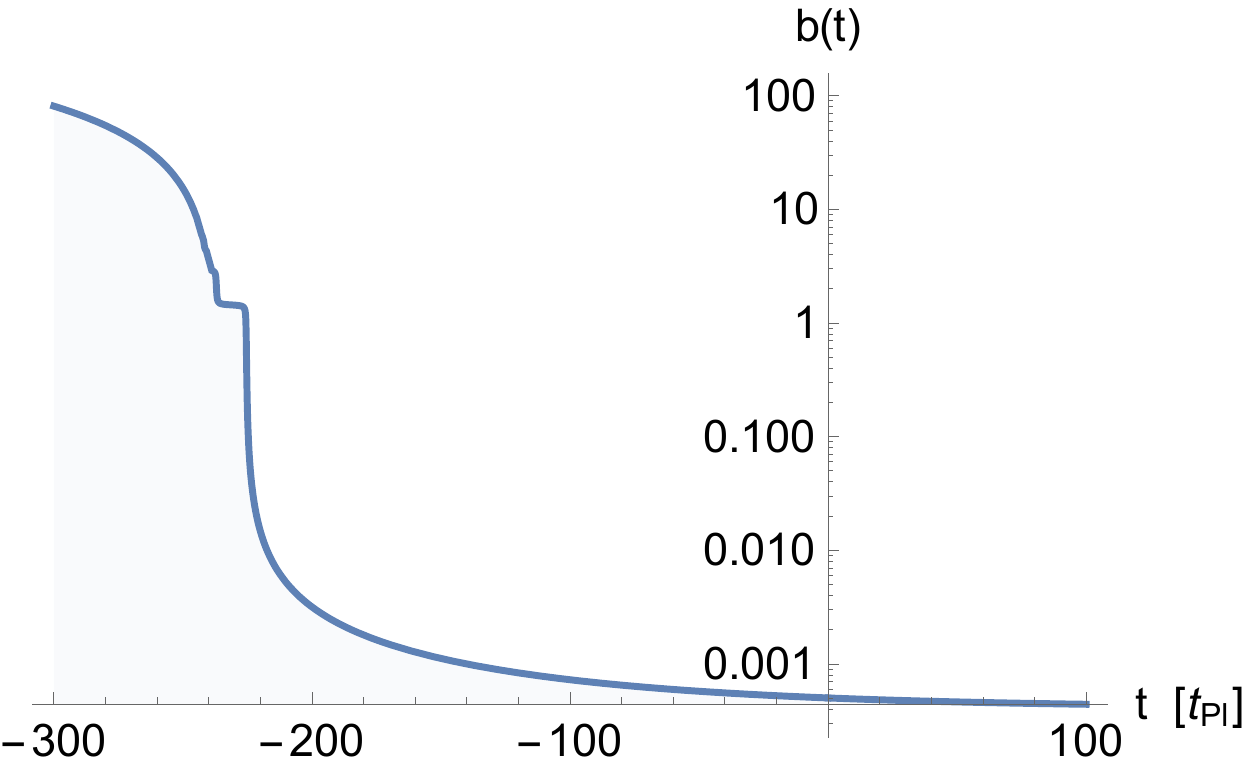}
\label{b vs t}
  \end{minipage}
  \vspace{0.2 cm}
  \begin{minipage}[t]{0.4\textwidth}
\includegraphics[scale=0.55]{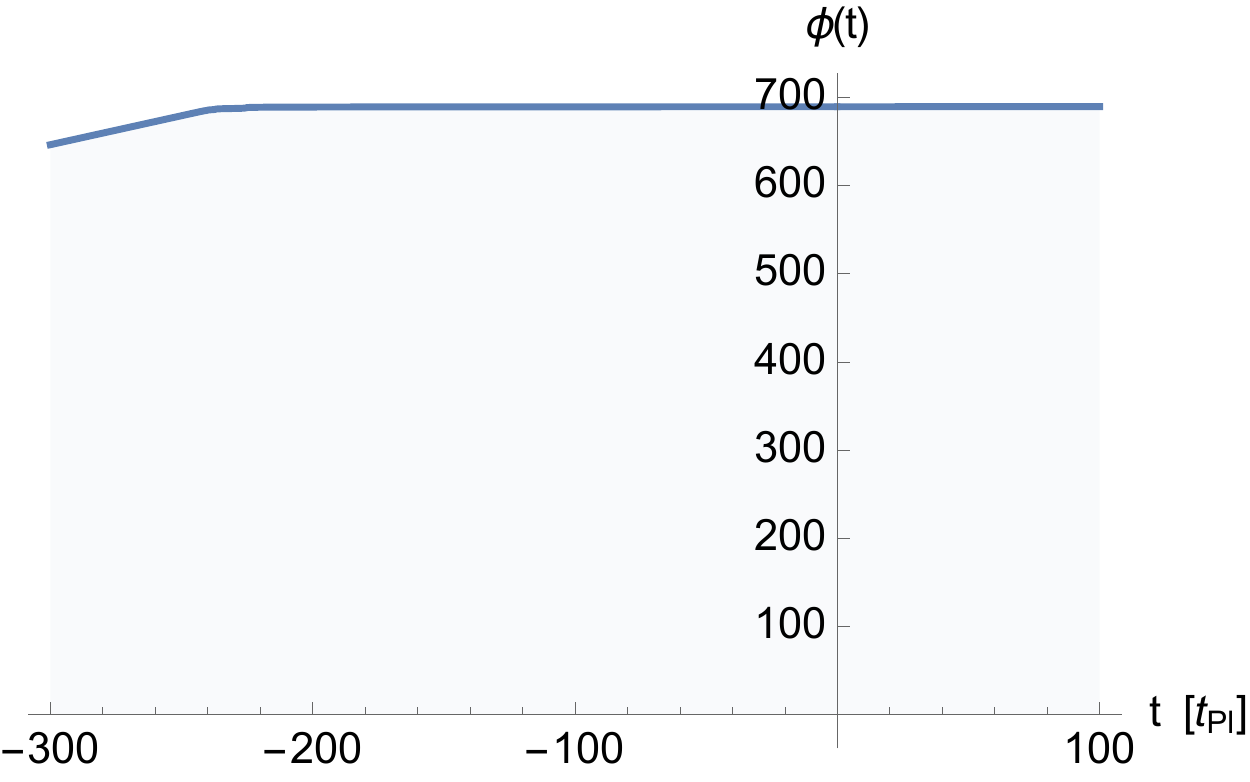}
\label{phi vs t}
  \end{minipage}
    \hfill
   \caption{\underline{Upper panel:} $b(t)$ forward evolution, starting from the (quasi)static phase at $t=-300$. \underline{Lower panel:} $\phi(t)$ forward evolution, also starting from the (quasi)static phase at $t=-300$.} 
   \label{forward evolutions}
\end{figure}

The forward evolution of the $b$ parameter which, together with $v$, characterizes the gravitational sector of the background dynamics, is shown in the upper panel of Fig. \ref{forward evolutions}. As expected, one can check that this parameter nearly vanishes in the classical regime. The lower panel shows the field evolution during the emergent phase and it can be noticed that, unlikely to what happens in the contracting branch of the usual LQC bounce (see {\it e.g.} \cite{bl}), the field does not oscillate. Instead, it remains almost constant. We have studied different field trajectories associated with many different initial conditions and, although  the field value does vary during the static phase, oscillations have never been observed.
The evolution of $b$ shows ``kinks" which start at times corresponding to a scale factor local minimum and last until the next minimum is reached. In LQC, only one minimum is present (at the big bounce). Anyway, both in LQC and QRLG, it just corresponds to phases where $b$ suddenly speeds up. Those phases connect an initial (postbounce) an final (prebounce) evolution during which $b$ is almost constant.\\ 

\begin{figure}[!tbp]
  \centering
  \hspace{-0.9 cm}
  \begin{minipage}[t]{0.4\textwidth}
 \includegraphics[scale=0.58]{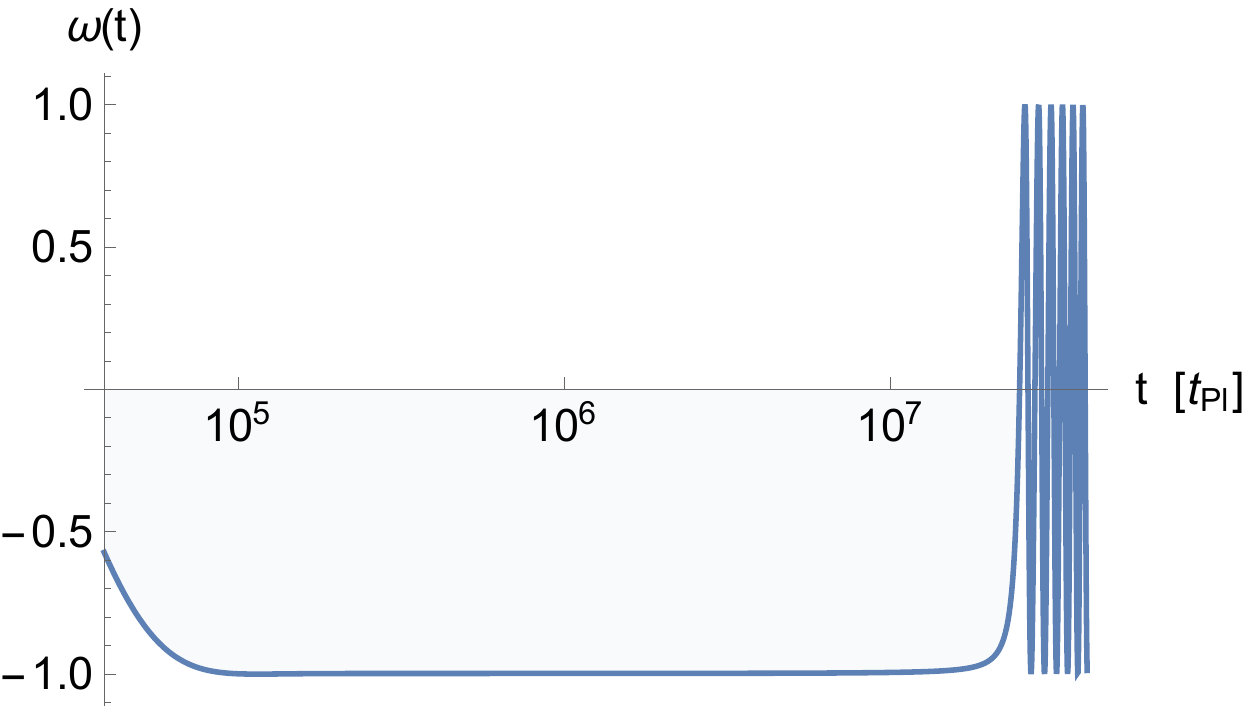}
\label{omega forward}
  \end{minipage}
  \vspace{0.2 cm}
  \begin{minipage}[t]{0.4\textwidth}
\includegraphics[scale=0.55]{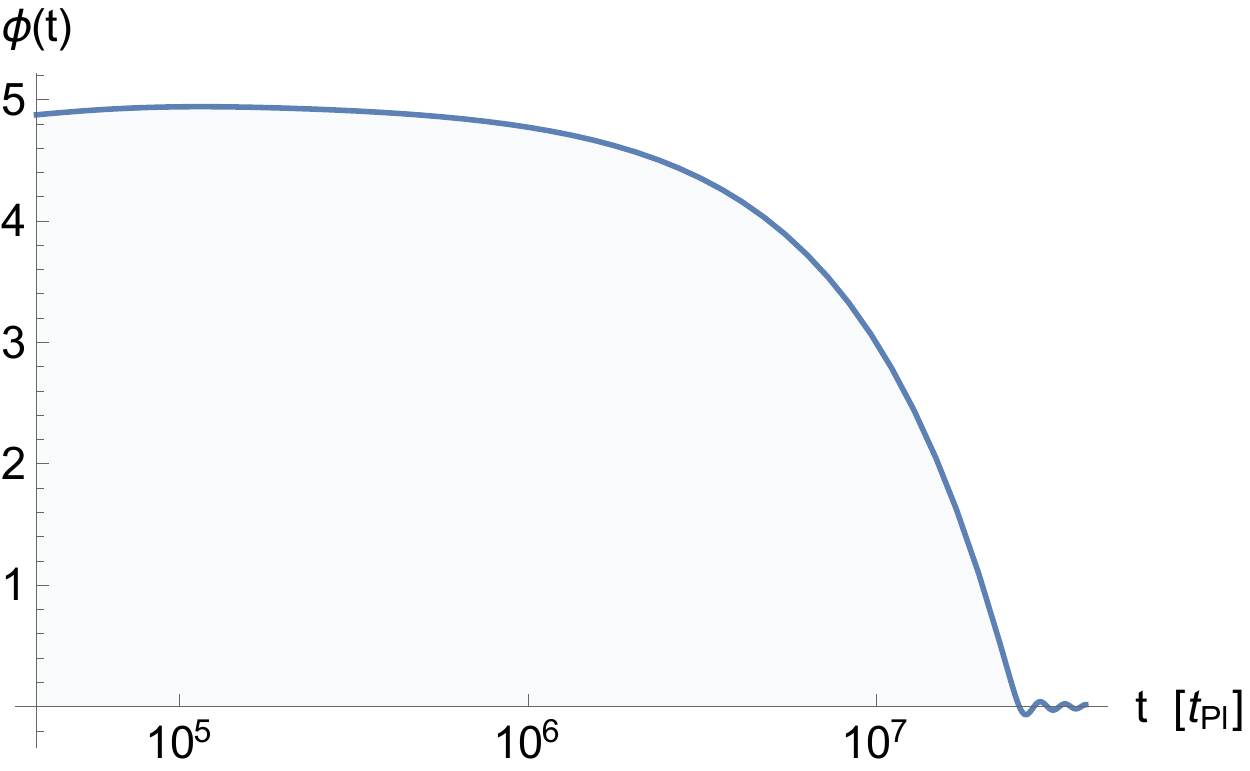}
\label{phi forward}
  \end{minipage}
   \caption{\underline{Upper panel:} $\omega(t)= \mathcal{P}(t)/\rho(t)$ evolution during the slow-roll phase. \underline{Lower panel:} Scalar field evolution during the slow-roll phase.} 
   \label{Inflation Omega Phi}
\end{figure}

As it is well known, inflation is a strong attractor once the correct matter content is set (see {\it e.g.} \cite{bl,Linsefors:2014tna, Martineau:2017sti,Ashtekar:2011rm} for recent results on this point in the framework of LQC). This is not a specific LQC feature but this comes as a result of the presence of a scalar field together with a high enough initial energy density \cite{Bolliet:2017czc,Barrau:2017rwl}. It is therefore no surprise that in QRLG too the static phase is generically followed by an inflationary stage, as can be seen in Fig. \ref{Inflation Omega Phi}.
In this figure $\omega$ is the dimensionless ratio between the scalar field pressure 

\begin{equation}
\mathcal{P}(t) = E_{\text{kin}}(t) - E_{\text{pot}}(t) = \frac{1}{2} \frac{P_{\phi}(t)^{2}}{\left(2 \pi \gamma v(t) \right)^{2}} - \frac{1}{2} m^{2} \phi(t)^{2} ~,
\end{equation}

and the scalar field energy density

\begin{equation}
\rho(t) = E_{\text{kin}}(t) + E_{\text{pot}}(t) = \frac{1}{2} \frac{P_{\phi}(t)^{2}}{\left(2 \pi \gamma v(t) \right)^{2}} + \frac{1}{2} m^{2} \phi(t)^{2} ~.
\label{Field Energy Density}
\end{equation}

It characterizes the cosmological perfect fluid equation of state. When $\omega \rightarrow -1$ the scalar field acts as a positive cosmological constant and generates inflation in a quasi-de Sitter stage. \\

The equation of state parameter evolution, presented in the upper panel of Fig. \ref{Inflation Omega Phi}, together with the field evolution presented in the lower panel of Fig. \ref{Inflation Omega Phi}, are typical of a slow-roll inflationary phase\footnote{On those plots the  initial field value $\phi_{\text{in}}=4$ has been chosen smaller than for the plots presented in Fig. \ref{forward evolutions} in order to make the figure easier to read. The $\omega$ and $\phi$ behaviors remain qualitatively equivalent, the only difference being the duration of inflation.}. The duration of the phase of slow-roll inflation has no influence on the shape of the primordial power spectra given as a function of the comoving wave number (as long as it lasts long enough to ensure the freezing of the considered modes). The number of inflationary e-folds is however of crucial importance to relate the computed primordial power spectra to cosmological microwave background (CMB) observations. This number determines the portion of the comoving spectrum which falls into the observational window.

\section{Inflation duration}

As stated previously, the knowledge of the number of inflationary e-folds is necessary as soon as one wants to compare the primordial power spectra with CMB observations: the position of the observed interval depends on this parameter. If the spectrum is fully scale invariant, this is of course nonrelevant -- this is why the total number of inflationary e-folds can be anything above 60-70 in usual cosmology -- but as soon as some specific features (like in LQC and QRLG) exist in the spectrum this is mandatory knowledge.\\

A  crude estimate of the observational window position is given by the comoving wave number associated with the size of the observable universe at the recombination time. The physical size of the observable universe at this time is of the order of (we still use Planck units)

\begin{equation}
L_{\text{rec}} \sim 4 \times 10^{58} ~.
\end{equation}

The associated physical wave number is therefore of the order of $k_{\varphi, \text{rec}} = 2\pi/ L_{\text{rec}} \sim 10^{-58}$.\\

To switch from physical coordinates to comoving ones, one needs to know the number of e-folds between the stationary state of the Universe and the recombination period, as the scale factor is normalized in the initial state (the chosen value is of course in itself arbitrary). This total number of e-folds can be expressed as the sum of the number of inflationary e-folds and the number of e-folds between the end of inflation and the recombination.
The number of e-folds of inflation can be expressed as

\begin{equation}
N = \ln \left(\frac{a(t_{e})}{a(t_{i})} \right) = \frac{1}{3} \ln \left( \frac{v(t_{e})}{v(t_{i})} \right) ~~,
\end{equation}

where $t_{i}$ and $t_{e}$ respectively correspond to the beginning and the end of the inflationary period. 
The number of e-folds between the end of the inflationary phase and recombination, denoted as $N'$, depends both on the well-known decoupling temperature (see, {\it e.g.} \cite{Kolb:1990vq}) and on the far less-constrained reheating one:

\begin{equation}
N' = \ln \left(\frac{T_{\text{rh}}}{T_{\text{dec}}} \right)   \simeq \ln \left(\frac{T_{\text{rh}}}{0.2~ \text{eV}} \right) \simeq 59,
\end{equation}

for a reheating temperature around the grand unification (GUT) scale (to fix the orders of magnitude).

The comoving wave number associated to the physical wave number $k_{\varphi, \text{rec}}=10^{-58}$ is therefore given by

\begin{equation}
k_{c, \text{rec}} = k_{\varphi, \text{rec}} a(t_{\text{rec}}) = k_{\varphi, \text{rec}} a(t_{i}) e^{59+N}.
\end{equation}

This makes the observational window dependence on the number of inflationary e-folds explicit. It can be noticed that when $N = 75$, $ k_{c,\text{rec}} \sim 1$ for $a(t_{i})=1$ (as usually chosen). This is why, as soon as the number of e-folds is substantially higher than the minimum required value, the part of the spectrum which is probed corresponds to modes with $k_{c}\gg 1$\\

In the following, we focus on the duration of inflation for different sets of initial conditions. Both for numerical convenience and because the stationary phase is, by definition, time translation invariant, the initial conditions are set just at the end of the static phase (corresponding to $t=-240$ in the simulation), before any significant growth of the physical volume. The situation is slightly more subtle when dealing with perturbations. Since we cannot explore fully the four-dimensional parameter space, we fix $b_{\text{in}}$ to the value obtained from the backward evolution. This is not an arbitrary choice -- unlike it would be for the matter content --, and this parameter does not enter the field energy density expression given in Eq. \ref{Field Energy Density}. Three variables are therefore remaining free: $\left\lbrace v_{\text{in}}, \phi_{\text{in}} , P_{\phi, \text{in}} \right\rbrace$. Since the relevant study for our purpose is the impact of both $v_{\text{in}}$ and $\phi_{\text{in}}$ on the duration of inflation, we set $P_{\phi, \text{in}}$ such that the initial Hamiltonian constraint is satisfied. \\

\begin{figure}[]
  \hspace{-1.2 cm}
  \begin{minipage}[t]{0.4\textwidth}
\includegraphics[scale=0.48]{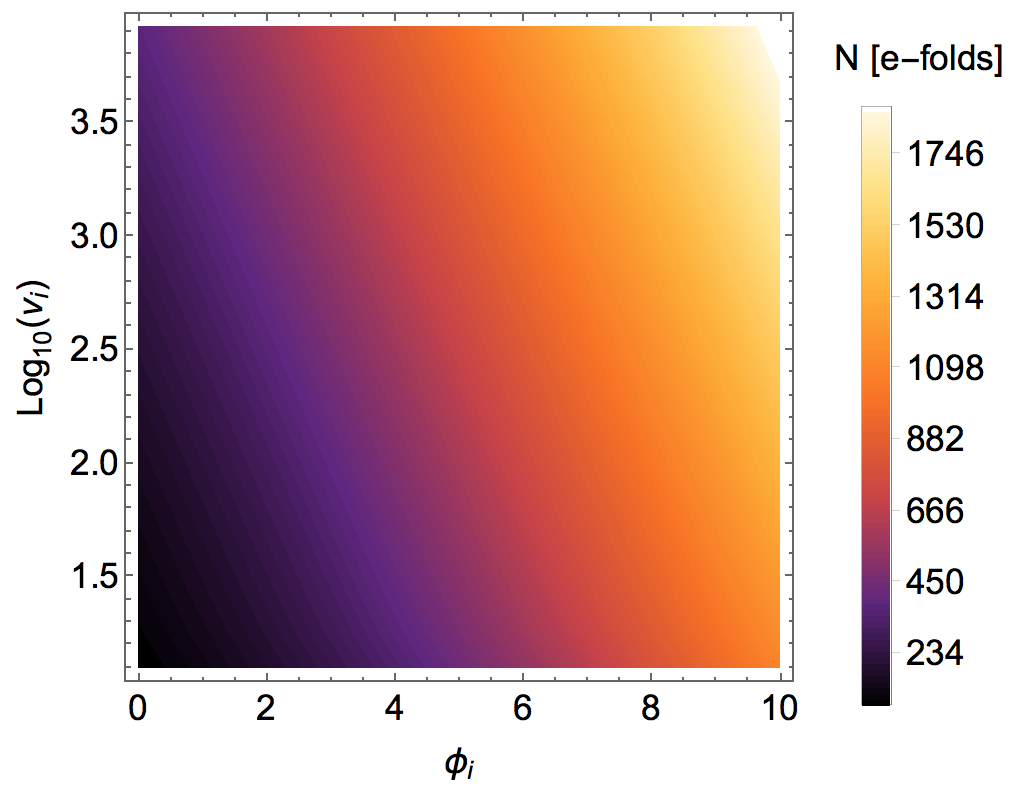}
  \end{minipage}
  \vspace{0.2 cm}
  \hspace{-1.2 cm}
  \begin{minipage}[t]{0.4\textwidth}
\includegraphics[scale=0.48]{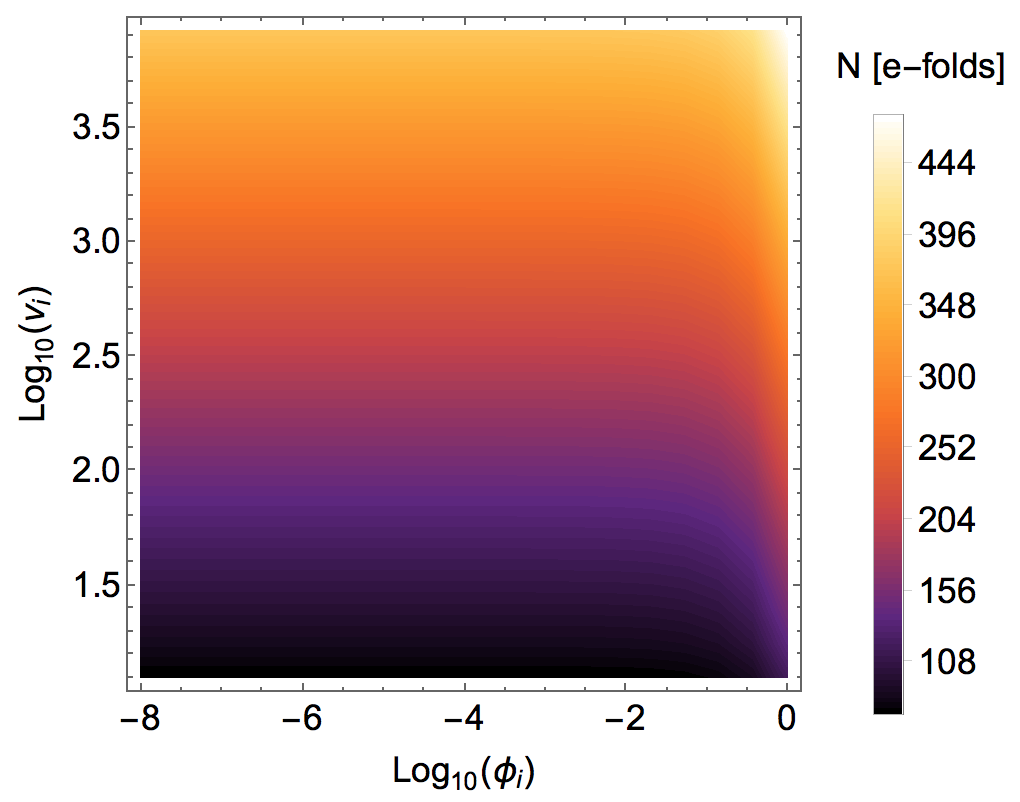}
  \end{minipage}
   \caption{\underline{Upper panel:} The number of inflationary e-folds $N$ as a function of $v_{\text{in}}$ and $\phi_{\text{in}}$. \underline{Lower panel:}  Zoom on the small values of $\phi_{\text{in}}$ in order to probe the low values of $N$.} 
   \label{E-folds slow roll vs v_in and Phi_in}
\end{figure}

The main results are displayed in Fig. \ref{E-folds slow roll vs v_in and Phi_in}. It can be seen that $N$ increases both with $v_{\text{in}}$ and $\phi_{\text{in}}$. The $\phi_{\text{in}}$ dependence is the same as in the LQC framework when initial conditions are set at the bounce. \\ 

The comparison with LQC is however subtle. In LQC, nearly any number of inflationary e-folds, including $N=N^{\star}\approx 60$ (which is interesting for phenomenology as this makes the nontrivial features observable) is possible if initial conditions are fine-tuned. If they are set at the bounce, there is no obvious preferred initial value for the field -- or alternatively for the sometimes used $x$ variable defined as the dimensionless square root of the potential energy density -- and it is hard to find a preferred inflation duration. The other way around, if initial conditions are set in the remote past of the contracting branch, and if the bounce energy density is fixed (the usual value being $\rho_{c} \approx 0.24$), a preferred value close to $N\approx 140$ does appear  \cite{bl,Linsefors:2014tna, Martineau:2017sti}.

In QRLG, the prediction of the number of e-folds is therefore similar to what happens in LQC, but only when initial conditions are set at the bounce, in the sense that the selection criterion for a preferred field value is lost. However, in QRLG the number of e-folds also depends on other parameters: $N$ clearly increases with $v_{\text{in}}$. \\

\begin{figure}[]
\hspace{-2.2 cm}
  \begin{minipage}[t]{0.4\textwidth}
\includegraphics[scale=0.46]{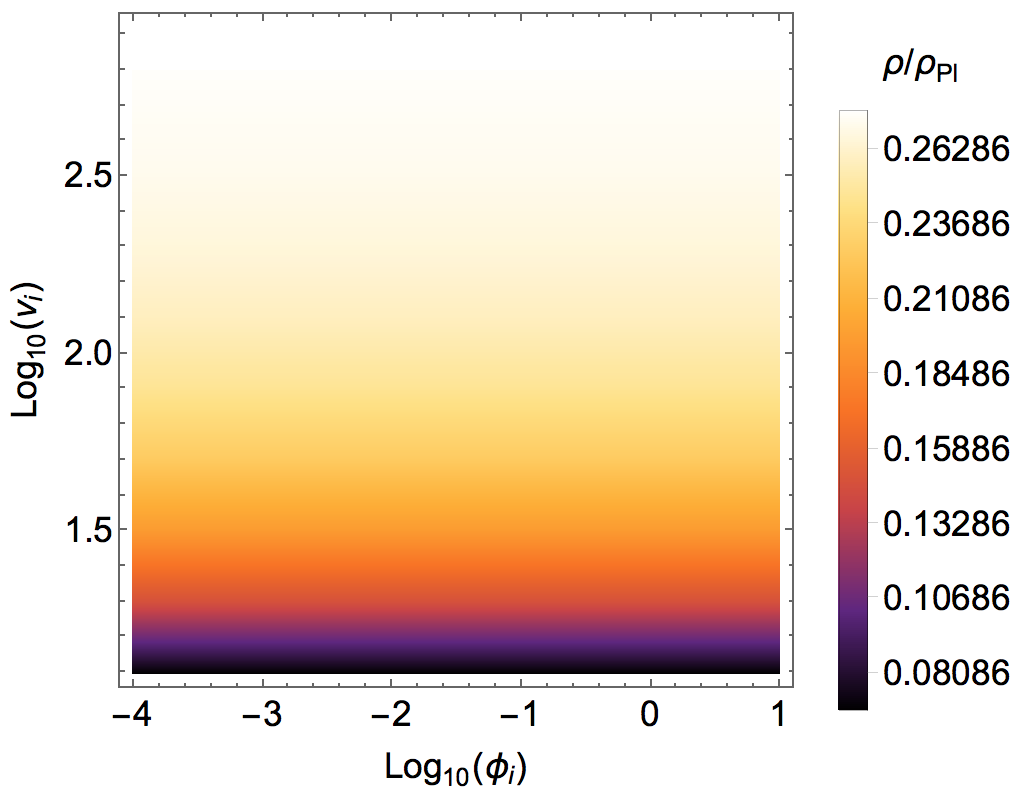}
  \end{minipage}
  \vspace{1 cm}
  \hspace{-2.2 cm}
  \begin{minipage}[t]{0.4\textwidth}
  \vspace{0.05 cm}
\includegraphics[scale=0.46]{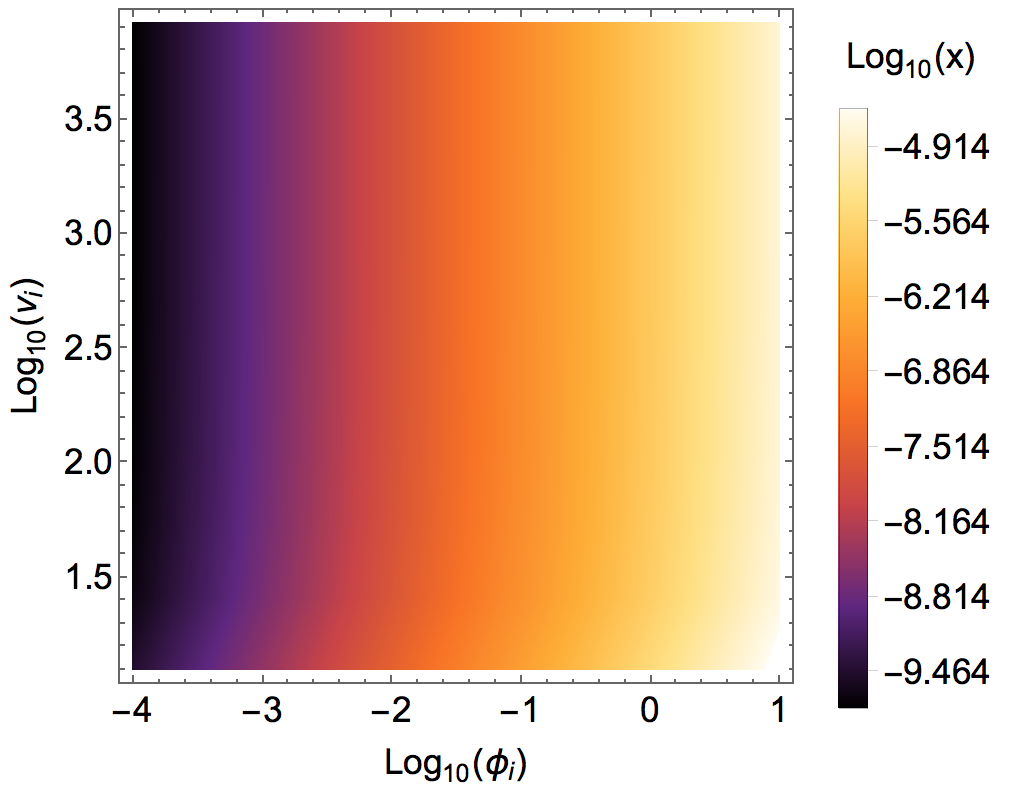}
  \end{minipage}
   \caption{\underline{Upper panel:} The field energy density at the end of the static phase as a function of $v_{\text{in}}$ and $\phi_{\text{in}}$. \underline{Lower panel:} The dimensionless ratio $x$ at the end of the static phase as a function of $v_{\text{in}}$ and $\phi_{\text{in}}$. } 
   \label{rho and x vs v_in and Phi_in}
\end{figure}

Numerical investigations show that, with this procedure, it is necessary to have $v_{\text{in}}> 12.4$, otherwise the Hamiltonian constraint cannot be fulfilled. This means that low values of $v_{\text{in}}$ are unaccessible, making a small number of e-folds, close to $N^{\star}$, even less probable than in LQC, not to say strictly impossible. The way $v_{\text{in}}$ does depend on $N$ may seem strange at first sight since the field energy density $\rho_{\text{in}} \propto v_{\text{in}}^{-2}$. In general the higher the density, the bigger the number of e-folds. However this effect is ``overcompensated'' by the fact that $\rho_{\text{in}}$ is also proportional to $P_{\phi,\text{in}}^{2}$ which increases when $v_{\text{in}}$ increases (in order to satisfy the Hamiltonian constraint). The initial field energy density $v_{\text{in}}$ dependence can be seen in Fig. \ref{rho and x vs v_in and Phi_in}.\\

For slow-roll inflationary models with a single inflaton field in the LQC framework, the number of e-folds depends on two parameters: the field energy density and the dimensionless ratio

\begin{equation}
x = \sqrt{\frac{E_{\text{pot}}(t)}{\rho(t)}} = \sqrt{ \frac{\frac{1}{2} m^{2} \phi(t)^{2}}{ \frac{1}{2} \frac{P_{\phi}(t)^{2}}{\left(2 \pi \gamma v(t) \right)^{2}} + \frac{1}{2} m^{2} \phi(t)^{2} }},
\end{equation}

at the beginning of the inflationary phase. It increases when those parameters increase. Basically $N \propto x^2$ and $N \propto \rho_{\text{in}}$ \cite{bl}. There is also a known field mass dependence $N \propto \ln \left( \frac{2}{m}\sqrt{\frac{\kappa}{3}\rho_{\text{in}}}\right)$. \\

In Fig. \ref{rho and x vs v_in and Phi_in}, we show the dependence of both $\rho$ and $x$ at the end of the stationary phase (\textit{i.e} at $t=-240$), upon initial conditions.  As previously mentioned, since the mass is set to $m=1.21\times10^{-6}$, the field energy density is kinetically dominated in the range of $\phi_{in}$ values presented here. The energy density therefore increases with $v_{in}$ but remains constant when $\phi_{in}$ varies. It would be possible to probe initial field values close to $10^{6}$ to study how $\rho$ varies when the potential term is no longer negligible. It is however not relevant to go into the details when $\phi_{in} \gg 10$ as the number of e-folds is in this case (and whatever the other parameters are) very high, as shown Fig. \ref{E-folds slow roll vs v_in and Phi_in}. On the other hand, this switches on the dependence for $x_{in}$. Since $\rho_{in}$ is almost constant with respect to variations of $v_{in}$, the value of $x_{in}$ only depends on the initial field value. If we consider together the $v_{in}$ and $\phi_{in}$ dependence of both $x_{in}$ and $\rho_{in}$ this leads to the trend which appears in the upper panel of Fig. \ref{E-folds slow roll vs v_in and Phi_in}. This confirms that the duration of inflation in QRLG, as in LQC, depends on the couple $\left \lbrace x, \rho \right\rbrace$ at the beginning of the inflationary period.\\

In summary, the duration of inflation in QRLG can be set close to the lower boundary $N^{\star} = 60$ but it requires a very high level of fine-tuning, even more important than in LQC when setting initial conditions at the bounce. For almost all the probed initial parameter space, the inflation duration is lengthy, pushing the observational window far in the ultraviolet part of the spectra. That is quite bad news for phenomenology as this makes the specific features of the model nearly impossible to observe. But this is good news for the consistency: the model agrees with observations (assuming that the observed tensor spectrum will be scale free) for nearly all its parameter space.\\

Now that the background dynamics has been defined and characterized, cosmological perturbations can be propagated on this background to derive the primordial power spectra. As precise calculations for perturbations in QRLG are still missing, we make here the hypothesis that perturbations are described by the usual theory.

\section{Primordial power spectra}

When dealing with a flat universe filled with a scalar field, the first-order perturbed Einstein equations are equivalent to the gauge-invariant Mukhanov-Sasaki equation:

\begin{equation}
\nu''(\eta, \vec{x}) - c_{s}^{2} \Delta \nu(\eta, \vec{x}) - \dfrac{z_{T/S}''(\eta)}{z_{T/S}(\eta)} \nu(\eta, \vec{x}) = 0 ~,
\label{Mukhanov}
\end{equation}

in which :
\begin{itemize}
\item $\nu$ is a gauge-invariant canonical variable built as a combination of the metric coordinate (Bardeen variables) and of the scalar field perturbations.
\item $z$ is the background variable that models the background impact on the perturbations and whose expression depends on the kind of inhomogeneities considered. The $T/S$ indices refer either to tensor or to scalar modes.
\item $c_{s}$ is the speed of sound, which is equal to the speed of light $c_{s}=1$ for a canonical scalar field.
\item The $^\prime$ symbol corresponds to a derivative with respect to the conformal time $\eta$.
\end{itemize}

As it can be seen from \eqref{Mukhanov}, the evolution of cosmological perturbations is equivalent to the one of a scalar field $\nu$ with a time dependent mass $m^{2} = - z_{T/S}''/z_{T/S}$ in a Minkowski space-time. Because of the dynamical background, the energy of the perturbations is not conserved (they can extract energy from the background evolution), hence the mass time dependence.\\

When quantizing the theory, the $\nu$ functions and their conjugate momenta become operators.  The associated Fourier temporal mode functions satisfy

\begin{equation}
\nu_{k}''(\eta) + \left(k_{c}^{2} - \dfrac{z_{T/S}''(\eta)}{z_{T/S}(\eta)} \right) \nu_{k}(\eta) =0 ~,
\label{MukhanovSasaki Temporal modes}
\end{equation}

in which $k_{c}$ corresponds to a comoving wave number.  

This equation can be recast in cosmic time :

\begin{equation}
    \begin{aligned}
& \ddot{\nu_{k}}(t) + H(t) \dot{\nu_{k}}(t)\\
&  + \left[ \left (\frac{k_{c}}{a} \right)^{2} - \dfrac{\dot{z}_{T/S}(t)}{z_{T/S}(t)}H(t) - \dfrac{\ddot{z}_{T/S}(t)}{z_{T/S}(t)} \right] \nu_{k}(t) =0.
\end{aligned}
  \label{MukhanovSasakiCosmicTime}
\end{equation}

We introduce a new parameter $h_{k}(t)=\nu_{k}(t)/a(t)$ such that \eqref{MukhanovSasakiCosmicTime} becomes

\begin{equation}
\begin{aligned}
& \ddot{h_{k}}(t)+3H(t)\dot{h_{k}}(t) + h_{k}(t) \\
& \left[H(t)^{2}+\dfrac{\ddot{a}(t)}{a(t)}+ \left(\frac{k_{c}}{a} \right)^{2} - H \dfrac{\dot{z}_{T/S}(t)}{z_{T/S}(t)} - \dfrac{\ddot{z}_{T/S}(t)}{z_{T/S}(t)} \right] =0.
\end{aligned}
\label{Mukhanov hk Cosmic Time}
\end{equation}

For the purpose of writing \eqref{Mukhanov hk Cosmic Time} as a set of two first order ordinary differential equations (ODE) we introduce a second parameter, $g_{k}(t) = a(t) \dot{h_{k}}(t)$, such that 

\begin{equation}
\left\{
    \begin{aligned}
& \dot{h_{k}}(t) = \dfrac{1}{a(t)} g_{k}(t)~, \\
& \dot{g_{k}}(t) = - 2 H(t) g_{k}(t) - a(t)h_{k}(t) \times \\
&  \left[ H(t)^{2} + \dfrac{\ddot{a}(t)}{a(t)} + \left (\frac{k_{c}}{a} \right)^{2} - H(t)\dfrac{\dot{z}_{T/S}(t)}{z_{T/S}(t)} - \dfrac{\ddot{z}_{T/S}(t)}{z_{T/S}(t)} \right]. 
\end{aligned}
  \right.
  \label{Set EDO Mukhanov}
\end{equation}

Finally the primordial power spectra are respectively defined by

\begin{equation}
\mathcal{P}_{T}(k_{c}) = \dfrac{4 \kappa k^{3}}{\pi^{2}} \left|\dfrac{\nu_{k}(t_{e})}{z_{T}(t_{e})} \right|^{2} 
\end{equation}

for tensor modes, and

\begin{equation}
\mathcal{P}_{S}(k_{c}) = \dfrac{k^{3}}{2 \pi^{2}} \left|\dfrac{\nu_{k}(t_{e})}{z_{S}(t_{e})} \right|^{2}
\end{equation}

for scalar ones, in which $t_{e}$ stands for the cosmic time at the end of the slow-roll phase.

Since the scale factor is deduced from $v(t)$ by

\begin{equation}
a(t) = \left( \frac{2 \pi \gamma v(t)}{V_{0}} \right)^{1/3} ~~,
\end{equation}

the value of $V_{0}$ will have an impact on the spectra. This dependence will be later discussed.\\

In this article we assume ``usual" perturbations on a QRLG background. This is obviously only a first step in the direction of a full QRLG treatment. The question of perturbations in LQC is a tricky one. On the one hand, the dressed metric \cite{Agullo1,Agullo2,Agullo3} (which is close to hybrid quantization \cite{Gomar:2015oea} from the observational viewpoint) puts the emphasis on the quantum aspects of both the background and the perturbations, while the deformed algebra \cite{Bojowald:2011aa,eucl3,eucl2} highlights the consistency and gauge aspects. Those issues will need to be dealt with in QRLG in the future.\\

Another point that needs to be addressed is the question of initial conditions for perturbations, which is a well-known and tricky one. Basically, the idea is to go far enough in the past so that the effective potential $z_{T/S}''/z_{T/S}$ is negligible compared to $k_{c}^{2}$ and the evolution equation becomes the one of a harmonic oscillator. This is the case in the de Sitter background of standard inflation. This is also the case in the contracting phase of the LQC bounce for tensor modes. A detailed discussion for the more complicated case of scalar perturbations in LQC can be found in \cite{Schander:2015eja} for the philosophy followed in this study. (Another approach based on the definition on a fourth order adiabatic vacuum at the bounce can be found in \cite{Agullo1,Agullo2,Agullo3}.)

In the following, the simulations presented rely on an initial state for perturbations defined in the Minkowski vacuum. The choice of the precise initial vacuum is however not crucial at this stage as the aim of the study is to investigate the way in which the spectra depend on the QRLG parameters. Considering different vacua generally induces only small modifications in this framework. This point has been investigated for LQC and it was shown that although the vacuum choice makes some differences in the IR, most of the features of the spectrum remain unchanged \cite{Barrau:2018gyz}. We have effectively tested different vacua and, as it could have been expected, the results and conclusions drawn below were checked not to depend on the precise choice.

It should be mentioned that several interesting features of the primordial power spectra have been derived in \cite{Brandenberger:2017pjz} for a background behavior similar to the model considered in this work. In particular, damped oscillation appears at scales smaller than a characteristic value and the reddening of the spectrum increases at all the scales when the number of small bounces increases.

\subsection{Primordial tensor power spectra}

For tensor perturbations, the background variable is given by $z_{T}(t) = a(t)$, and the previous set of ODEs \eqref{Set EDO Mukhanov} becomes:

\begin{equation}
\left\{
    \begin{aligned}
& \dot{h_{k}}(t) = \dfrac{1}{a(t)} g_{k}(t)~, \\
& \dot{g_{k}}(t) = - 2 H(t) g_{k}(t) - \frac{k_{c}^{2}}{a(t)} h_{k}(t) ~.
\end{aligned}
  \right.
  \label{Set EDO tensor modes RG}
\end{equation}

\begin{figure}[]
\hspace{-0.8 cm}
  \begin{minipage}[t]{0.4\textwidth}
\includegraphics[scale=0.435]{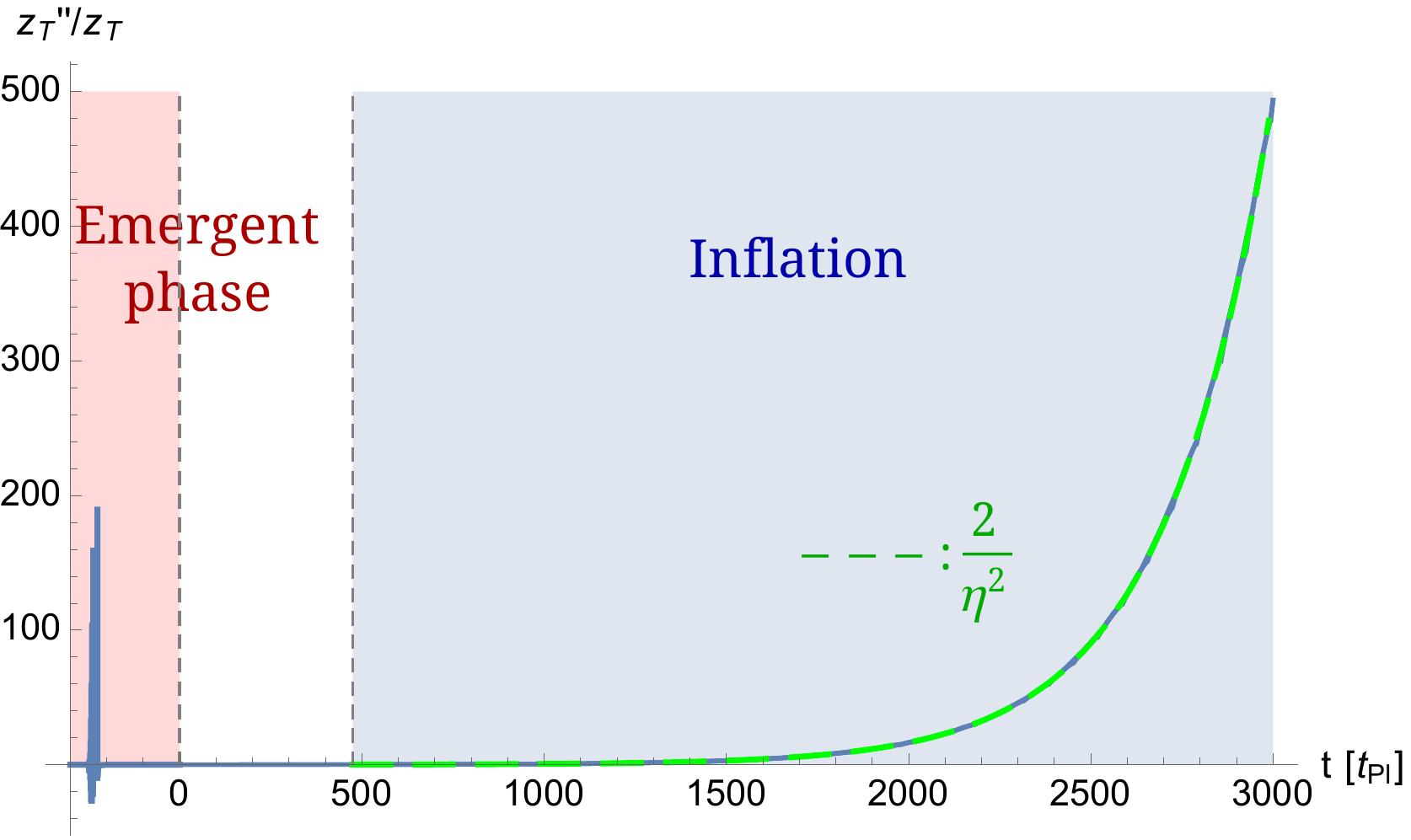}
  \end{minipage}
  \vspace{0.2 cm}
    \hspace{-1 cm}
  \begin{minipage}[t]{0.4\textwidth}
\includegraphics[scale=0.55]{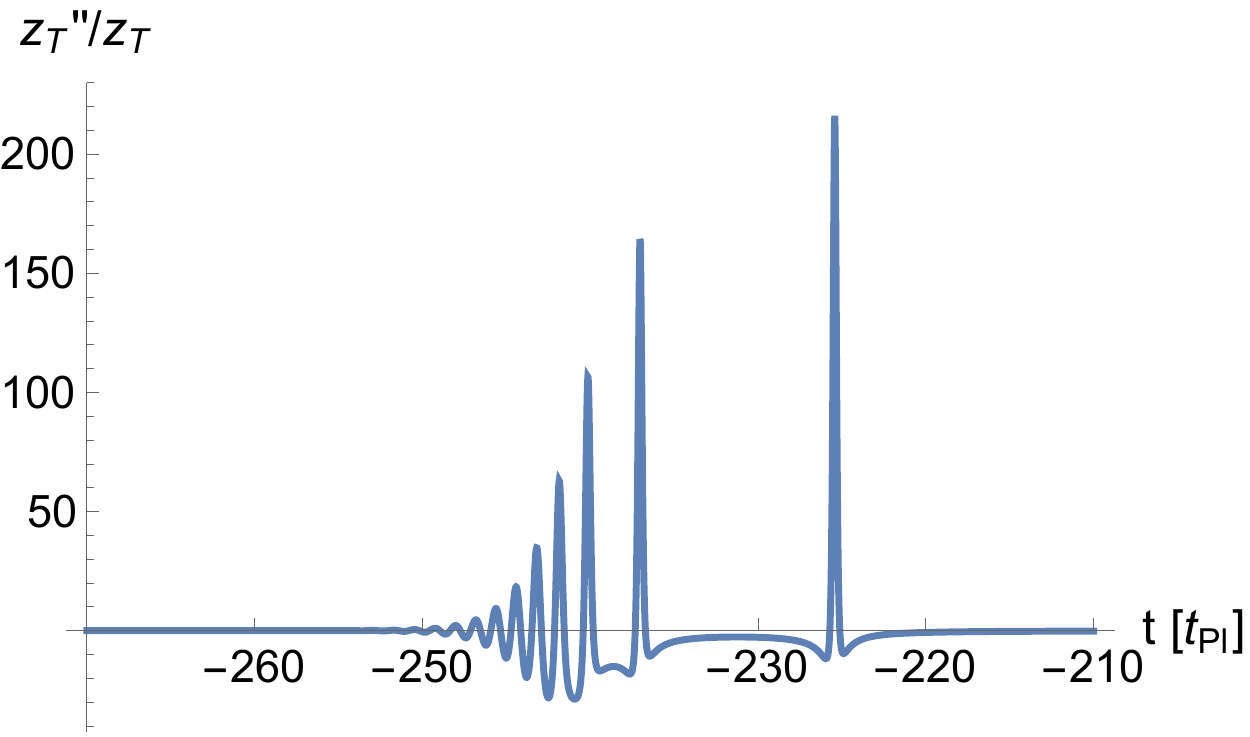}
  \end{minipage}
   \caption{The tensor potential $z_{T}''/z_{T}$. \underline{Upper panel:} Full range including the beginning of the inflationary phase. \underline{Lower panel:} Zoom on the emergent phase.} 
   \label{Tensor potential}
\end{figure}

What matters for the shape of primordial spectra is the tensor potential $z_{T}''(t)/z_{T}(t)=a''(t)/a(t)$. More precisely, the key point is the relative value of the potential and of the comoving wave number $k_{c}$. The evolution of this potential in QRLG is presented in Fig. \ref{Tensor potential}.\\

The evolution of the potential should be slightly contrasted with what happens in LQC when initial conditions are set in the classical contracting branch. In usual LQC, the tensor potential converges quickly toward zero and a Bunch-Davies vacuum can be properly defined for all modes as long as initial conditions are set sufficiently far away from the bounce. This initial normalization, combined with the potential behavior $z_{T}''/z_{T} \simeq 2/\eta^{2}$ during the slow-roll phase, leads to a scale invariant spectrum for all modes of cosmological interest\footnote{It should be notices that this only holds for tensor modes, as it is impossible to properly define a Bunch-Davies vacuum in the same way for scalar perturbations.}. In addition, the bounce leads to a peak in the potential that creates oscillation in the intermediate part of the spectrum \cite{Mielczarek:2010bh,Bolliet:2015bka,Agullo:2015tca}. In the specific case of the deformed algebra approach a UV divergence also occurs, due to an effective change of signature of the metrics \cite{lcbg}, but this situation will not be considered here.\\

In QRLG, the tensor potential also exhibits a $2/\eta^{2}$ evolution during the inflationary phase, as it can be seen in the upper panel of Fig. \ref{Tensor potential}. When the simulation is started, the potential is of the order of $z_{T}''(t_{i})/z_{T}(t_{i}) \simeq 10^{-5}$, for $V_{0}=1$. It is, at this stage, not easy to analytically demonstrate that the potential strictly vanishes in the remote past. Thus, the choice of the initial state as a vacuum (in the sense $k_{c}^{2} \gg z_{T}''(t_{i})/z_{T}(t_{i})$) is no more physical for comoving wavenumbers that do not satisfy $k_{c}^{2} \gg 10^{-5}$. This means that the IR limit of the spectrum might not be fully reliable and deserves future investigation. This is however not important for phenomenology as the observational window anyway falls in the intermediate or UV part. As in standard LQC, the bounces induce peaks in the potential leading to oscillations in the spectra.\\

\begin{figure}[!tbp]
  \centering
  \begin{minipage}[t]{0.4\textwidth}
  \hspace{0.5 cm}
\includegraphics[scale=0.35]{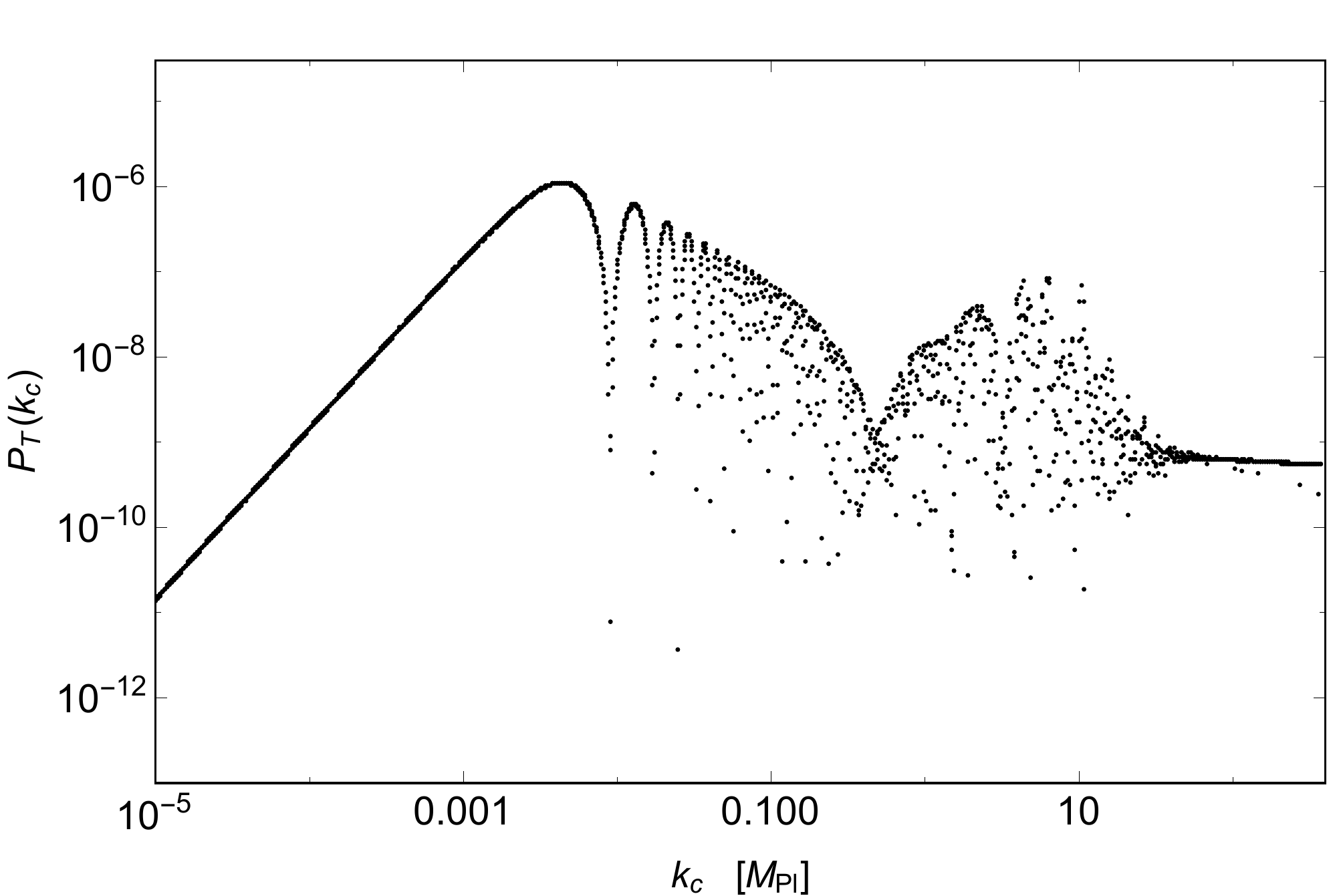}
\label{b vs t}
  \end{minipage}
  \vspace{0.2 cm}
  \begin{minipage}[t]{0.4\textwidth}
\includegraphics[scale=0.35]{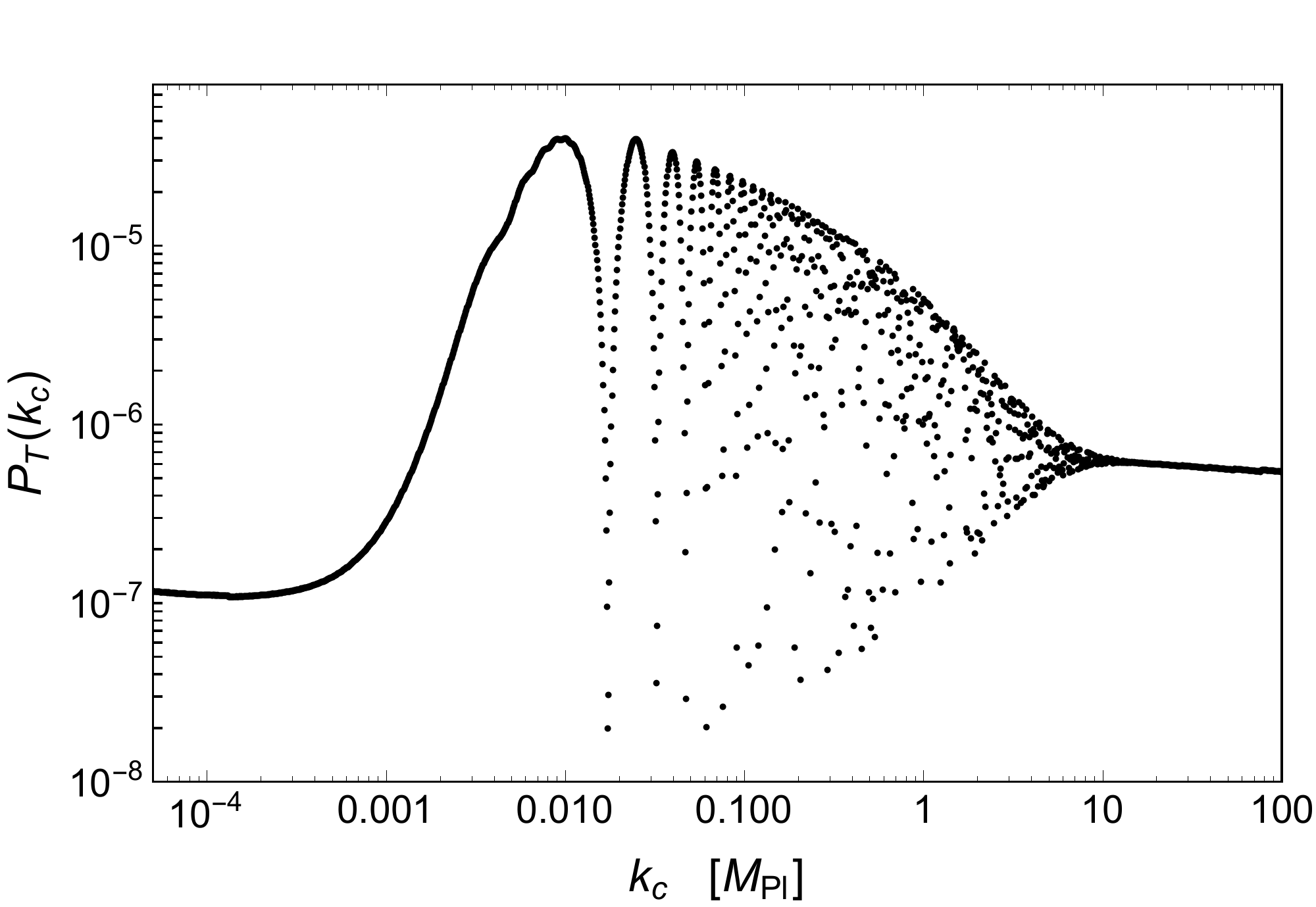}
\label{phi vs t}
  \end{minipage}
    \hfill
   \caption{\underline{Upper panel:} Primordial tensor power spectrum of cosmological perturbations on a QRLG background, with  $V_{0}=1$ and $\phi_{in} =4$. \underline{Lower panel:} Primordial tensor power spectrum of cosmological perturbations on a LQC background (with a larger inflaton mass).} 
   \label{spectrum V0=1}
\end{figure}

The primordial tensor power spectrum is represented in the upper panel of Fig. \ref{spectrum V0=1}, with arbitrarily chosen values $V_{0}= 1$ and $\phi_{in}= 4$. The spectrum dependence upon those parameters will be discussed in the
following paragraphs. As expected, one can notice a rising IR part, a scale-invariant UV part (corresponding to $k_{c} > 30$ in this case) and an intermediate oscillatory part – with a richer structure than in usual LQC due to the multiple minibounces. 
For an easy comparison, the typical LQC spectrum is shown on the lower panel of \ref{spectrum V0=1} (the difference in amplitude is just due to a different mass value chosen for numerical convenience).

Following the study of the previous section on the duration of inflation, it can be concluded that for almost all the parameter space of initial conditions, the observable part of the QRLG primordial power spectrum is nearly scale invariant, as in GR. Probing deviations with respect to GR, that is the oscillatory intermediate regime, the initial conditions have to be highly fine-tuned so that $N$ approaches $N^{\star}$.\\

\begin{figure}[!h]
\begin{center}
\hspace{-1.6 cm}
\includegraphics[scale=0.33]{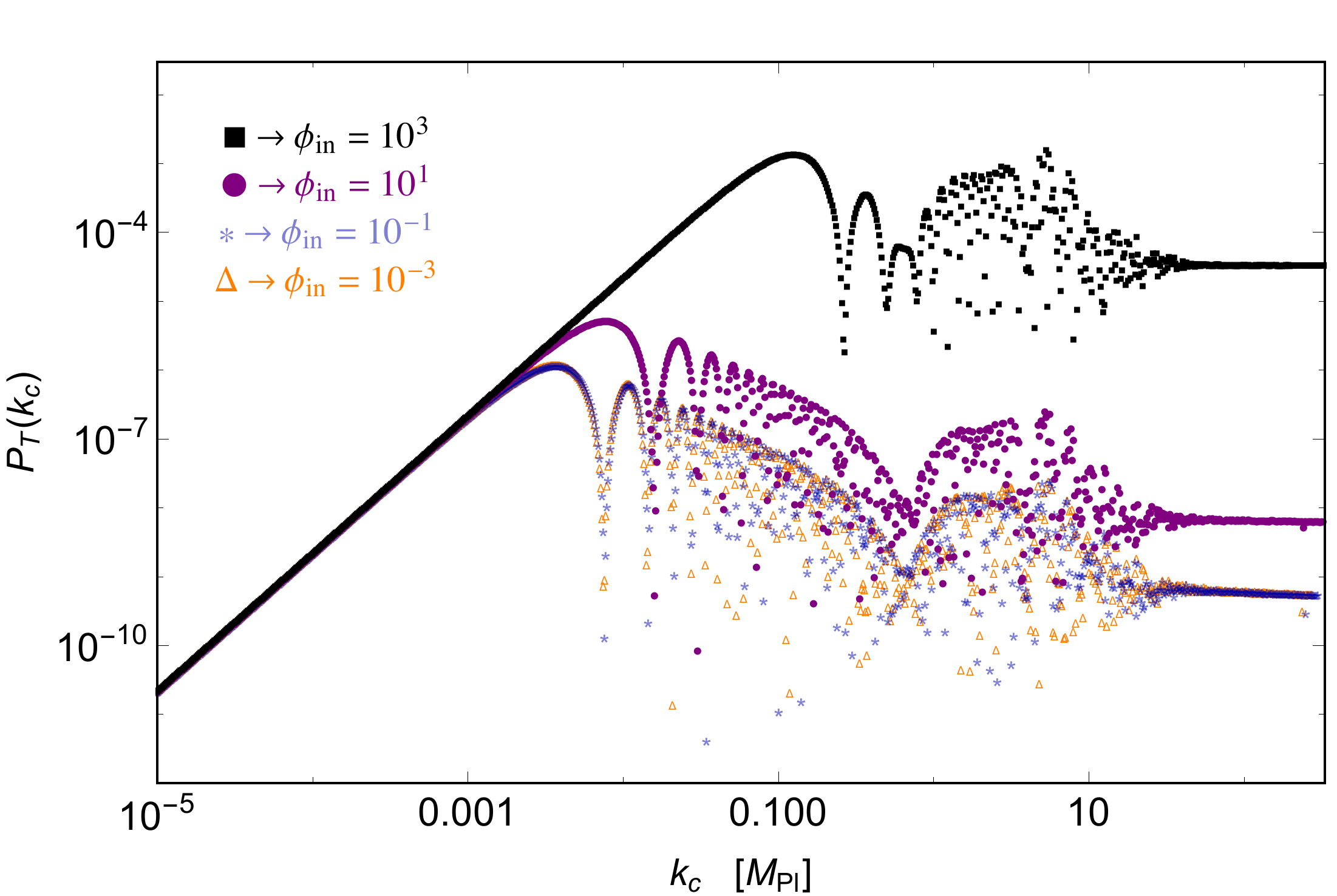}
\caption{Primordial tensor power spectrum for different initial values of the scalar field $\phi_{in}$ in the static phase and $V_{0}=1$ .} 
\label{spectra vs Phi}
\end{center}
\end{figure}

In Fig. \ref{spectra vs Phi}, we show the impact of the initial value of the scalar field on the spectra. For the initial field values chosen here, $\left\lbrace 10^{-3},10^{-1},10^{1},10^{3} \right\rbrace$, the field energy density -- which is the physical parameter -- is (due to the low mass of the field) fully kinetic energy dominated and equal to $0.24$. The spectrum amplitude increases with the initial field value, and reaches a level which is in disagreement with observations when the initial value is  $\gtrsim 10$,  taking into account the upper bound on the tensor to scalar ratio \cite{PCP2015}. This sets a bound on possible initial field values : $\phi_{in} \lesssim10$. It is interesting to notice that spectra with $\phi_{in}= \left\lbrace 10^{-3},10^{-1}\right\rbrace$ perfectly overlap. The field value is mostly irrelevant, regarding the tensor spectra, as soon as it is  $\lesssim 0.1$. 

\begin{figure}[!h]
\begin{center}
\includegraphics[scale=0.33]{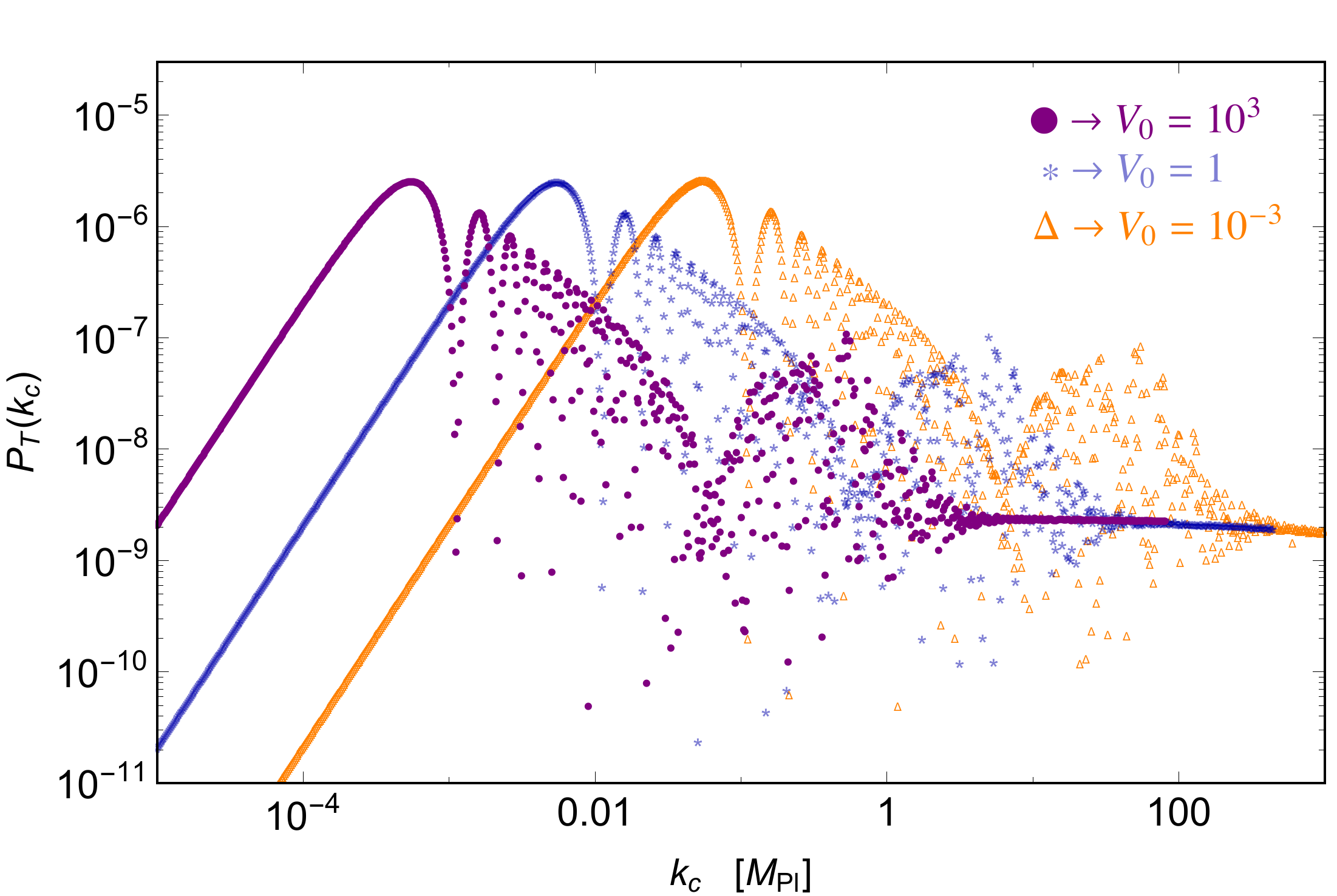}
\caption{Primordial tensor power spectra of cosmological perturbations on a QRLG background for different values of $V_{0}$.} 
\label{tensor spectra different V0}
\end{center}
\end{figure}

In Fig. \ref{tensor spectra different V0}, the impact of different choices for  $V_{0}$ are shown. Since the tensor potential writes 

\begin{eqnarray}
\frac{z_{T}''(t)}{z_{T}(t)} &=& a(t) \ddot{a}(t) + \dot{a}(t)^{2} \\ \nonumber
&=& \left(\frac{2 \pi \gamma}{V_{0}} \right)^{2/3} \left[ v(t)^{1/3} \ddot{v(t)^{1/3}}  + \left( \dot{v(t)^{1/3}} \right) ^{2}\right]  \\ \nonumber
&\propto & V_{0}^{-2/3},
\end{eqnarray}

and as  this value should be compared to the squared comoving wave number, the horizontal shift of the spectra, proportional to $V_{0}^{-1/3}$, can easily be anticipated. This agrees with the numerical results. \\

It is therefore in principle possible to constrain $V_{0}$ by requiring the appropriate properties of the spectrum in the observable window. Once $N$ is fixed the observable window position, given by $k_{c, \text{rec}}$, depends only on $a_{0}$, and thus on $V_{0}^{1/3}$ if $v_{in}$ is fixed. As the tensor power spectrum has not yet been measured and as there is still a degeneracy with the number of e-folds this is only a prospective claim at this stage.

\subsection{Primordial scalar power spectra}

This study focuses on tensor modes. Scalar perturbations are more closely related to available observation but are substantially more difficult to deal with. In the case of scalar perturbations, the usual background variable is $z_{S}(t)=a(t)\dfrac{\dot{\phi}(t)}{H(t)}$ and Eqs. \eqref{Set EDO Mukhanov} cannot be solved analytically. Because of the more complex shape of the scalar potential, the fate of scalar perturbations is not as clear as for the tensor ones\footnote{This is already the case in usual LQC where the definition of a proper vacuum state is tricky.}.\\

\begin{figure}[!tbp]
  \begin{minipage}[t]{0.4\textwidth}
\includegraphics[scale=0.55]{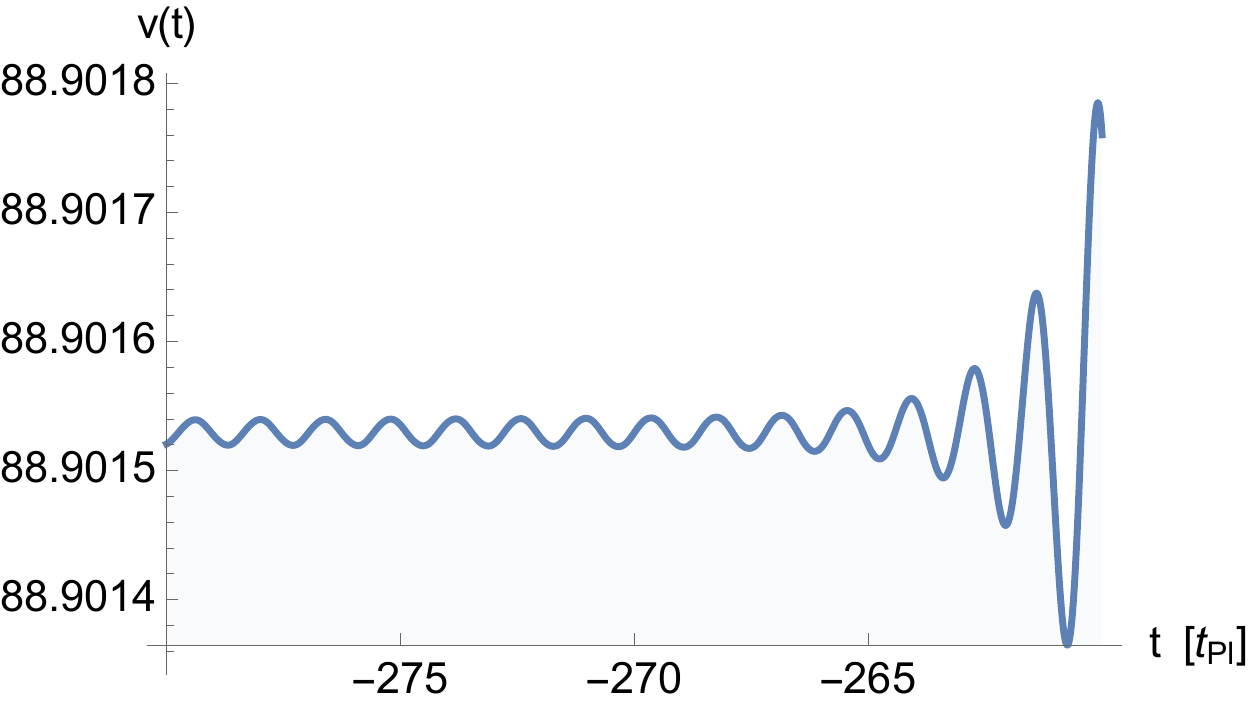}
  \end{minipage}
  \vspace{0.2 cm}
  \hspace{-0.3 cm}
  \begin{minipage}[t]{0.35\textwidth}
\includegraphics[scale=0.55]{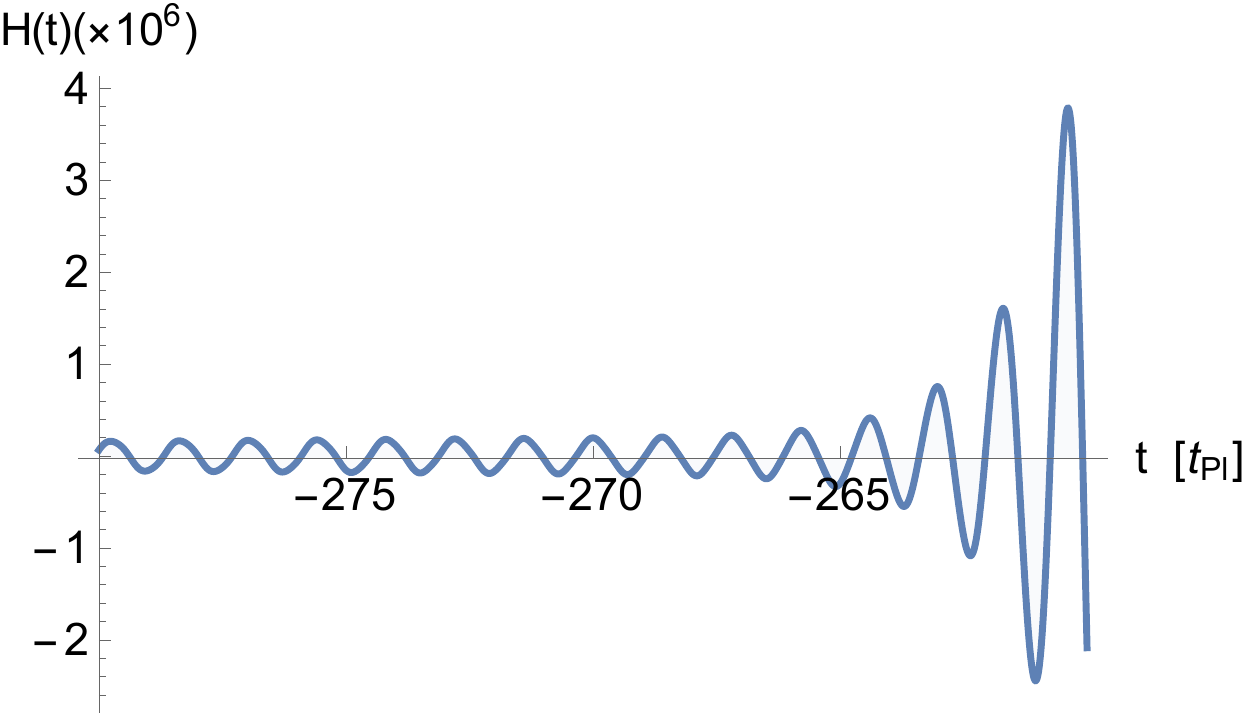}
  \end{minipage}
      \vspace{0.2 cm}
  \begin{minipage}[t]{0.48\textwidth}
\includegraphics[scale=0.55]{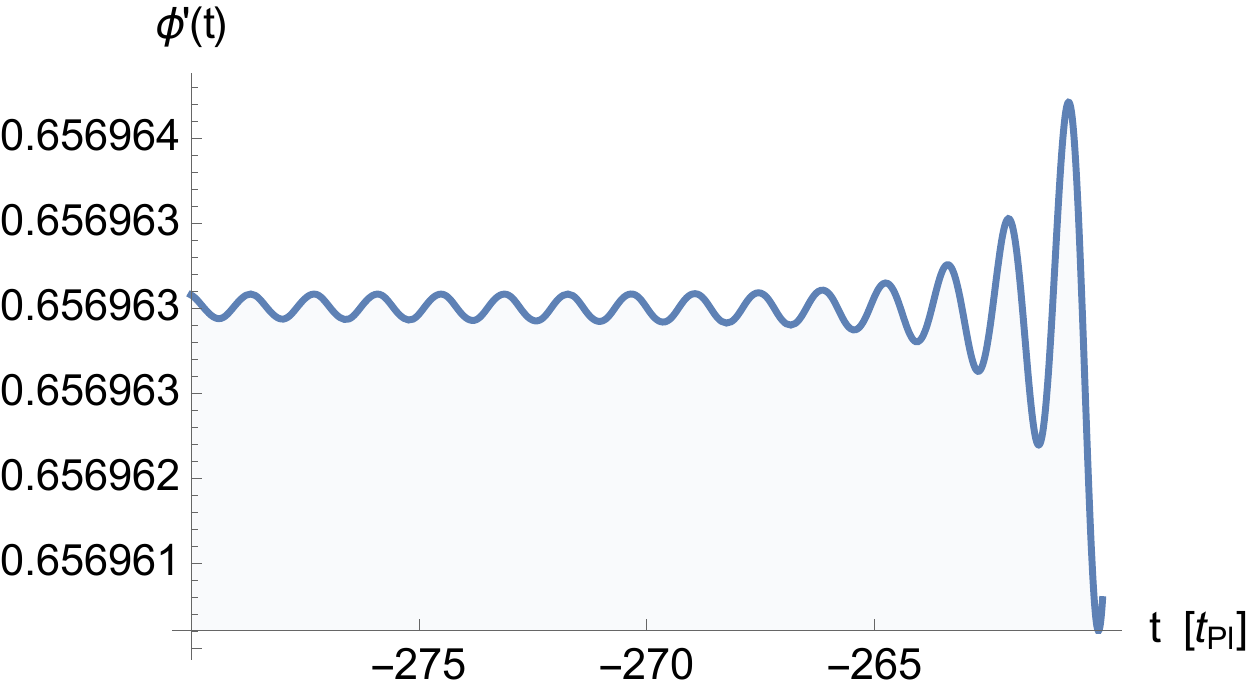}
  \end{minipage}
   \caption{Oscillatory behavior of the different parameters that compose the background scalar variable $z_{S}$ during the static phase. \underline{Upper panel:} background variable v(t). \underline{Middle panel:} Hubble parameter. \underline{Lower panel:} Time derivative of the field.} 
   \label{Scalar potential parameters plot}
\end{figure}

Due to the oscillating behavior of $v(t)$ in the static phase, as it can be seen in the upper panel of Fig. \ref{Scalar potential parameters plot}, the Hubble parameter $H(t)$ oscillates around 0, as shown in the middle panel of Fig. \ref{Scalar potential parameters plot}. This induces a nontrivial behavior of  $z_{S}(t)$. The scalar potential in the static phase therefore exhibits very fast oscillations of small amplitude that will amplify scalar perturbations. The resulting power spectrum is not physical. This is however not a clear conclusion as:

\begin{itemize}
\item It is probable that oscillations are actually damped when going far enough in the past, making the choice of a nonambiguous initial vacuum possible. Currently available simulations do not, however, allow to answer unambiguously this question because the case of scalar perturbations is quite intricate. The knowledge of the scale factor behavior is not sufficient and one needs the full constraint. The difficulty is however purely numerical and should be solved in a near future.
\item The Mukhanov-Sasaki variables might be modified in QRLG and their usual expressions might not hold anymore. The fact that ``usual'' perturbations are propagated is of course a very heavy hypothesis of this study. Building a fully self-consistent QRLG perturbation theory is a huge task.
\end{itemize}

\section{Influence of the field mass}

\begin{figure}[!tbp]
\hspace{-2.0 cm}
  \begin{minipage}[t]{0.4\textwidth}
\includegraphics[scale=0.38]{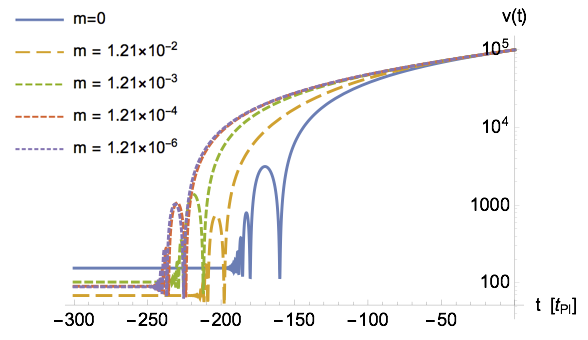}
  \end{minipage}
  \vspace{0.2 cm}
  \hspace{-2.0 cm}
  \begin{minipage}[t]{0.35\textwidth}
\includegraphics[scale=0.34]{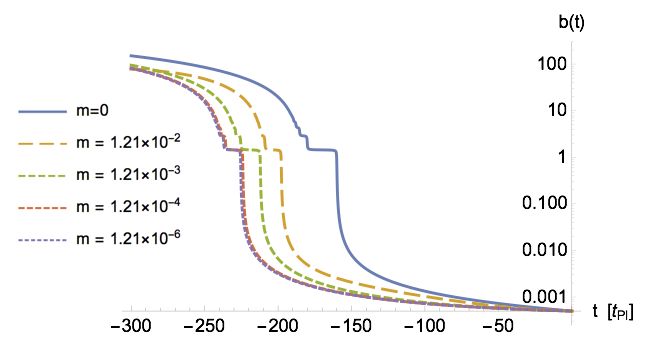}
  \end{minipage}
   \caption{\underline{Upper panel:} $v(t)$ backward evolution for different values of the scalar field mass, starting from $t=0$. \underline{Lower panel:} $b(t)$ backward evolution for different values of the scalar field mass, starting from $t=0$. } 
   \label{v and b different masses}
\end{figure}

The QRLG cosmological sector is very different from usual cosmology. Thus, the relevance  of the usual mass field value $m=1.21\times10^{-6}$ can be challenged. In the following, we consider different field masses around the usual one: $m=1.21\times10^{n}$, with $n=\left\lbrace-10, -8,-6,-4,-3,-2 \right \rbrace$. Changing the value of $n$ results in a small shift of the emergent background dynamics, which has no phenomenological importance, see Fig. \ref{v and b different masses}. In this figure the trajectories for all masses $m<1.21\times 10^{-6}$ perfectly overlap, both for $v$ and $b$.\\

However, a modification of the scalar field mass also modifies the postemergent dynamics, and notably the inflationary period.

\begin{figure}[!h]
\begin{center}
\hspace{-0.5 cm}
\includegraphics[scale=0.35]{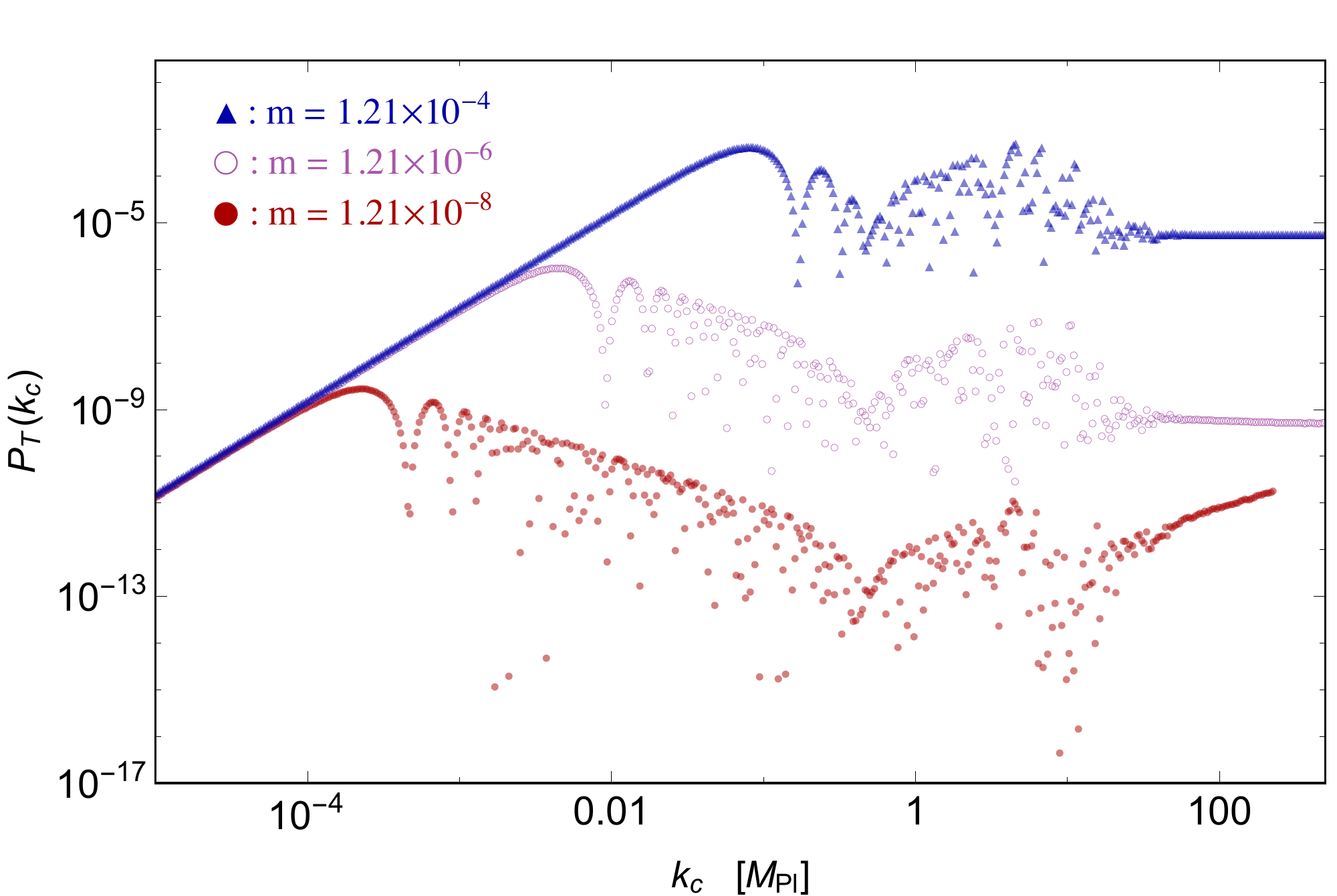}
\caption{Primordial tensor power spectra with $V_{0}=1$, $\phi_{in}=4$, and different field masses given by $n=\left\lbrace -8,-6,-4 \right \rbrace$.} 
\label{tensor spectra different m}
\end{center}
\end{figure}

The primordial tensor spectra for $n= \left\lbrace -8,-6,-4 \right\rbrace$ are represented Fig. \ref{tensor spectra different m}. It appears that, if initial conditions are chosen such that $\left\lbrace v_{in},b_{in},\phi_{in} \right\rbrace$ are fixed, and $P_{\phi,in}$ varies according to the Hamiltonian constraint, the general trend of the spectra, namely the rising IR behavior, the oscillations in the intermediate regime, and the scale invariance in the UV, do not depend on the mass. Different masses only result in a shift of the spectra, mostly as in LQC \cite{Bolliet:2015bka}. However, numerical simulations suggest that for extremely low values of the mass (typically $m<10^{-7}$), the shape of the power spectrum, even in the UV, is not scale-invariant anymore. This might be used as a constraint for the parameters of the model but this anyway requires a deeper treatment of the perturbations in QRLG.\\

If initial conditions are set such that $\left\lbrace v_{in},b_{in},P_{\phi,in} \right\rbrace$ are fixed and the initial field $\phi_{in}$ varies according to the constraint, then $\phi_{in}$ is lower by a factor $10^{n+6}$ with respect to the usual case $n=-6$. For example, if the field mass is close to the Planck mass, such as $n=-2$, then $\phi_{in}$ is divided by $10^{4}$. Those small values of $\phi_{in}$, together with the increase of the potential steepness with $m$,  lead to very small numbers of inflationary e-folds. The combination of those two effects can even prevent the slow-roll phase from happening. The associated tensor spectra are deeply modified, with a non-scale-invariant behavior presumably excluded by future observations.\\

The mass dependence of both the number of e-folds $N$ and of the tensor spectra shape highly depends on the way one deals with the Hamiltonian constraint. Variations of the mass can either induce a simple shift of the spectra or deeply modify the previously studied behavior. But substantial modifications only appear when large deviations (at least by 2 orders of magnitudes) from the usual value $m=1.21\times10^{-6}$ are considered. For reasonable deviations, all the conclusions previously stated still hold.

We also recall that the conclusions mentioned here are probably no longer true for scalar perturbations as the background variable $z_{S}$ directly depends on the field. The mass term therefore has a direct impact on the scalar potential, thus on the spectra, and may have a deeper effect than the simple shift observed in Fig. \ref{tensor spectra different m}.

\section{Conclusion}

Quantum reduced loop gravity is an important step in trying to bridge that gap between full quantum gravity and effective quantum cosmology. As the kinematics is defined before the minisuperspace reduction, some important features of LQG, such as the graph structure and SU(2) quantum numbers, are preserved although simplified to make relevant calculations analytically tractable. The key point is to impose the gauge-fixing conditions to the diagonal spatial metric. The minisuperspace reduction is then implemented at the dynamical level, keeping terms preserving the diagonality conditions.\\

The main result of QRLG is the replacement of the usual LQC bounce by an ``emergent + bounce'' scenario. In this article we have studied the inflationary dynamics in this framework. The main result when scanning the full parameter space is that the number of inflationary e-folds is always greater that the experimental lower bound around 60-70. This not fully true in LQC where the number of e-folds can be tuned to an arbitrary small number by choosing appropriate initial conditions. \\

We have also calculated the tensor power spectrum and shown its dependence upon the parameters of the model. The IR part is blue, the intermediate part is oscillatory and and the UV part is nearly scale invariant. Following the study on the number of e-folds, the observational window falls on the UV part and a flat tensor power spectrum is therefore predicted.\\

Several improvements are possible for future studies:
\begin{itemize}
\item The scalar power spectrum should be better investigated. This requires a deeper understanding of the background behavior in the remote past of the static phase.  
\item It might be possible to assign a known probability distribution function to a parameter driving the dynamics (like in \cite{bl,Martineau:2017sti}), but this requires a ``harmonic oscillator-like'' behavior in the deep past and this is not established at this stage.
\item Cosmological perturbations should also be QRLG-corrected.
\item The anisotropic version of QRLG should be investigated as the shear is expected to be possibly important at the bounce.
\end{itemize}

Refining quantum cosmological models is a major challenge. Although quite a lot of subtleties do appear when going from LQC or QRLG or group field theory (GFT) \cite{Gielen:2017eco,Gerhardt:2018byq}, it is interesting that the main global features seem to be preserved, making the overall picture more and more reliable.

\section*{Acknowledgments}

This work was supported in part by the NSF grant PHY-1505411, and the Eberly research funds of Penn State. 

Killian Martineau is supported by a grant from the C.F.M foundation.

\bibliography{refs}

\end{document}